\documentclass[lettersize,journal]{IEEEtran}
\usepackage{amsmath,amsfonts,amssymb,amsthm}
\usepackage{array}
\usepackage{textcomp}
\usepackage{stfloats}
\usepackage{url}
\usepackage{verbatim}
\usepackage{graphicx}

\def\BibTeX{{\rm B\kern-.05em{\sc i\kern-.025em b}\kern-.08em
    T\kern-.1667em\lower.7ex\hbox{E}\kern-.125emX}}
\usepackage{balance}
\usepackage{cite}
\usepackage[hidelinks]{hyperref}
\usepackage{xcolor}
\usepackage{multirow}
\usepackage{bm}

\usepackage{color}

\usepackage{caption}
\usepackage{subcaption}
\usepackage[labelformat=simple]{subcaption}

\usepackage[font=small]{caption}
\usepackage{pdfpages}

\usepackage{tikz}
\usetikzlibrary{tikzmark,decorations.pathreplacing}
\usepackage{rotating}

\newtheorem{theorem}{Theorem}

\newtheorem{lemma}{Lemma}
\newtheorem*{proof*}{Proof}
\newtheorem*{remark*}{Remark}
\newtheorem{remark}{Remark}

\usepackage{algorithm}
\usepackage[end]{algpseudocode}

\algnewcommand\algorithmicforeach{\textbf{for each}}
\algdef{S}[FOR]{ForEach}[1]{\algorithmicforeach\ #1\ \algorithmicdo}

\DeclareMathOperator*{\argmin}{arg\,min}

\usepackage{fancyhdr}

\usepackage{stfloats}

\begin{document}
\title{Hybrid Precoder Design for Angle-of-Departure Estimation with Limited-Resolution Phase Shifters}
\author{Huiping Huang, \textit{Member, IEEE}, Musa Furkan Keskin, \textit{Member, IEEE}, Henk Wymeersch, \textit{Fellow, IEEE}, \\
Xuesong Cai, \textit{Senior Member, IEEE}, Linlong Wu, \textit{Member, IEEE}, Johan Thunberg, \\
Fredrik Tufvesson, \textit{Fellow, IEEE}  \vspace{-4mm}
\thanks{This paper is partially supported by the Vinnova B5GPOS Project under Grant 2022-01640, HORIZON-MSCA NEAT-6G Project under Grant 101152670, and the Swedish Research Council grant 2022-03007, and is partially supported by the Wallenberg AI, Autonomous Systems and Software Program (WASP) funded by the Knut and Alice Wallenberg Foundation.}
\thanks{H. Huang, M. F. Keskin, and H. Wymeersch are with Department of Electrical Engineering, Chalmers University of Technology, 41296 Gothenburg, Sweden (e-mail: \{huiping; furkan; henkw\}@chalmers.se).}
\thanks{X. Cai, J. Thunberg, and F. Tufvesson are with Department of Electrical and Information Technology, Lund University, 22100 Lund, Sweden (e-mail: \{xuesong.cai; johan.thunberg; fredrik.tufvesson\}@eit.lth.se).}
\thanks{L. Wu is with the Interdisciplinary Centre for Security, Reliability and Trust (SnT), University of Luxembourg, 1855 Luxembourg, Luxembourg (e-mail: linlong.wu@uni.lu).}
}

\markboth{}
{Hybrid BF with Limited-resolution Phase Shifter}

\maketitle

\begin{abstract}
Hybrid analog-digital beamforming stands out as a  key enabler for future communication systems with a massive number of antennas. In this paper, we investigate the hybrid precoder design problem for angle-of-departure (AoD) estimation, where we take into account the practical constraint on the limited resolution of phase shifters. Our goal is to design a radio-frequency (RF) precoder and a base-band (BB) precoder to estimate AoD of the user with high accuracy. To this end, we propose a two-step strategy where we first obtain the fully digital precoder that minimizes the angle error bound, and then the resulting digital precoder is decomposed into an RF precoder and a BB precoder, based on the alternating optimization and the alternating direction method of multipliers. Furthermore, we derive the quantization error upper bound and provide convergence conditions for the proposed algorithm. Numerical results demonstrate the superior performance of the proposed method over state-of-the-art baselines.
\end{abstract}


\section{Introduction}
\label{introduction}

Millimeter wave (mmWave) and terahertz (THz) band have been proven to play an important role in future wireless systems, because they can provide ultra-high data rates \cite{Heath2016, Ge2023, Weng2023, Nazari2023, Chen2022, Chen2019, Petrov2020, Cai2022}. However, high carrier frequencies result in severe path loss. Large-scale antenna systems, which are equipped with hundreds or even thousands of antennas, have emerged as a crucial technology for addressing this problem \cite{Studer2016, Jacobsson2017, Bjornson2019, Keskin2023, Wei2021Nov}.

It is not feasible for large-scale antenna systems to employ fully digital beamforming at mmWave/THz, since fully digital beamforming requires as many radio-frequency (RF) chains (including digital-to-analog converters, mixers, etc.) as the antennas, leading to prohibitive hardware costs and power consumption \cite{Gao2015}. On the contrary, hybrid beamforming where only a small number of RF chains are needed is a promising solution to handle this problem \cite{Molisch2017}. The RF chains are connected to antennas via phase shifters with a finite number of quantized phases \cite{Li2020Jan, Lyu2021, Chen2017Aug, Wang2018May, Uwaechia2019, Li2020}.

Numerous works have been devoted to hybrid beamformer (precoder and/or combiner) design with practical constraints \cite{Zhang2005, Ayach2014, Wang2018July, Lyu2021, Dong2019, Zhang2019, Zhang2014, Yu2016, Gao2016, Sohrabi2015, Sohrabi2016, Chen2017Aug, Lin2017, Wang2018May, Uwaechia2019, Li2020}. Among them, the following four methods attract much attention. \textit{(i)} The authors in \cite{Lyu2021} proposed a hybrid beamforming algorithm with 1-bit resolution  phase shifter, which is based on alternating optimization framework and the Babai algorithm \cite{Babai1986} (termed as ``Alt-Babai''). \textit{(ii)} An iterative hybrid transceiver design approach using alternating optimization and coordinate descent method (CDM) was developed in \cite{Chen2017Aug} (termed as ``Alt-CDM''). \textit{(iii)} \cite{Ayach2014} exploited the spatial structure of mmWave channels and proposed a method for optimal unconstrained precoders and combiners, which employs sparse representation and orthogonal matching pursuit (termed as ``Spa-OMP''). \textit{(iv)} Another hybrid precoding method was presented in \cite{Yu2016}, which is on the basis of the manifold optimization \cite{Boumal2023, Absil2008, Hu2020} (termed as ``ManiOpt''). ManiOpt is a powerful technique when solving for a variable subject to Riemannian manifold constraints (such as unit-modulus, low-rank, etc.). However, the performance comes at the price of computational complexity and sensitivity to initializations.

Note that all the above-mentioned hybrid beamforming design methods are from the communications perspective.  In contrast, much less work has focused on hybrid beamforming design with limited-resolution phase shifters dedicated for channel parameters (such as angles, delays, Dopplers, etc.) estimation and positioning. Only \cite{Lin2021, Rajamaki2020, Cheng2023Aug} exactly relate to this topic. However, the results in \cite{Rajamaki2020, Cheng2023Aug} are presented from the sensing perspective. Moreover, \cite{Rajamaki2020} focuses on sparse array design. Several other related works considered hybrid beamforming design for integrated sensing and communications or target detection \cite{Wang2022Oct, Zhao2021, Hui2023, Liyanaarachchi2024, Cheng2023}. Note that the hybrid structures in \cite{Lin2021, Wang2022Oct, Zhao2021, Hui2023, Liyanaarachchi2024} are either \textit{partially-connected} or \textit{arbitrarily-connected}, resulting in a block-diagonal or block-sparse RF beamformer matrix, and their developed methods are heavily based on this feature. For instance, in \cite{Lin2021} the base-band (BB) beamforming matrix can be removed from the Fisher information matrix (FIM) due to block-diagonal structure, leading to a simpler Cram\'{e}r-Rao bound (CRB) matrix only related to the RF beamforming matrix. However, no existing works consider hybrid beamforming with \textit{fully-connected} structure for channel parameters estimation and positioning. Hybrid beamformer design with fully-connected structure is more challenging for the reason that more unknowns in the hybrid beamformer need to be estimated and no specific structure can be exploited. On the other hand, since fully-connected hybrid beamforming has the potential to achieve the full beamforming gain for each RF chain \cite{Zhao2021}, it is of great importance and has attracted much attention in recent years, see e.g., \cite{Chen2017Aug, Wang2018May, Uwaechia2019}. Note that the existing methods proposed in \cite{Lin2021, Wang2022Oct, Zhao2021, Hui2023, Liyanaarachchi2024} are not straightforwardly applicable to the fully-connected hybrid beamforming design, because the fully-connected hybrid beamformer matrix does not have block-diagonal or block-sparse structure. On the other hand, although optimal beamforming design for positioning has been investigated in e.g., \cite{Garcia2018, Keskin2022, Li2008, Deng2022}, these works investigated fully digital beamforming rather than hybrid beamforming. Note that existing hybrid beamformer design methods in \cite{Li2020Jan, Zhang2005, Ayach2014, Wang2018July, Lyu2021, Dong2019, Zhang2019, Zhang2014, Yu2016, Gao2016, Sohrabi2015, Sohrabi2016, Chen2017Aug, Lin2017, Wang2018May, Uwaechia2019, Li2020} can be applied for positioning. However, these methods do not guarantee a good performance in positioning (since they are proposed for the purpose of communications). Table \ref{tab:summary_hybridBF} summarizes the research domains, hybrid connection structures, and array geometries of recent works in hybrid beamforming. Even though \cite{Rajamaki2020} considers fully-connected hybrid beamforming for sensing and positioning, it limits the work to sparse arrays. Therefore, dedicated \textit{fully-connected hybrid beamforming design} for the purpose of \textit{sensing and positioning with arbitrary array geometry} remains unexplored in the existing literature.

\begin{table*}[t]
\caption{Summary of research domains, hybrid connection structures, and array geometries of recent works in hybrid beamforming}
\centering
\label{tab:summary_hybridBF}
\begin{tabular}{ |c|c|c|c|c|c|c|c|c|  }
\hline
\multirow{2}{4.4em}{References} & \multicolumn{3}{c|}{\textbf{Research domains}} & \multicolumn{3}{c|}{\textbf{Hybrid connection structures}} & \multicolumn{2}{c|}{\textbf{Array geometries}} \\ \cline{2-9}
& Communications & Sensing & Positioning & Partially & Arbitrarily & Fully & Dense & Sparse \\ 
\hline
\cite{Li2020Jan, Li2020} & \checkmark & & & & \checkmark & & \checkmark &  \\
\hline
\cite{Lyu2021} & \checkmark & & & \checkmark & \checkmark & & \checkmark &  \\
\hline
\multirow{3}{4.4em}{\cite{Chen2017Aug, Wang2018May, Uwaechia2019, Zhang2005, Ayach2014, Wang2018July, Dong2019, Zhang2019, Zhang2014, Sohrabi2015, Sohrabi2016, Lin2017}} & & & & & & & &  \\
 & \checkmark & & & & & \checkmark & \checkmark & \\
 & & & & & & & & \\
\hline
\cite{Lin2021} & & \checkmark & \checkmark & \checkmark & & & \checkmark &  \\
\hline
\cite{Rajamaki2020} & & \checkmark & \checkmark & & & \checkmark & & \checkmark \\
\hline
\cite{Cheng2023Aug} & & \checkmark & & & & \checkmark & \checkmark & \\
\hline
\cite{Wang2022Oct} & \checkmark & \checkmark & \checkmark & \checkmark & & & \checkmark &  \\
\hline
\cite{Zhao2021} & \checkmark & \checkmark & \checkmark & \checkmark & & & \checkmark & \\
\hline
\cite{Hui2023} & \checkmark & \checkmark & \checkmark & & \checkmark & & \checkmark &  \\
\hline
\cite{Liyanaarachchi2024} & \checkmark & \checkmark & \checkmark & \checkmark & & & \checkmark &  \\
\hline
\cite{Cheng2023} & & \checkmark & & \checkmark & & & \checkmark &  \\
\hline
This work & & \checkmark & \checkmark & & & \checkmark & \checkmark & \checkmark  \\
\hline
\end{tabular}
\end{table*}

Accurate estimation of angle-of-departure (AoD) and angle-of-arrival is crucial for improving user positioning accuracy, optimizing network efficiency, and enabling advanced communication technologies \cite{Garcia2018}. AoD estimation enhances location-based services, emergency response accuracy, and navigation systems by precisely determining the transmission direction of signals from base stations to user devices. Beamforming is crucial for optimizing spectrum use and improving signal quality in the future wireless networks, especially in the context of mmWave communications \cite{Kutty2016}. The precise estimation of AoD allows for the implementation of Internet of things, vehicle-to-everything communications, and smart city infrastructures by offering accurate positioning essential for smooth device connectivity. It enhances user experiences in applications like indoor navigation and targeted advertising by providing personalized information depending on the user's exact location. The importance of AoD in modern wireless communication systems highlights its function in fulfilling the need for fast, low-latency communications and in facilitating the growing variety of services and applications that rely on precise user placement \cite{Garcia2017}.

Therefore, in this paper we delve into the intricate problem of \textit{fully-connected} hybrid precoder design for AoD estimation, accounting for practical limitations on the \textit{finite resolution of phase shifters}. Our objective is to derive a solution comprising an RF precoder and a BB precoder that not only adheres to this practical constraint but also facilitates precise user AoD estimation. To achieve this goal, we follow the standard two-step strategy. That is, we first find a fully digital precoder that minimizes the angle error bound, which is the theoretical lower bound on AoD estimation. Then, we decompose the resulting digital precoder into an RF precoder and a BB precoder, by using alternating optimization framework and the alternating direction method of multipliers (ADMM). The numerical results show that the proposed method outperforms existing state-of-the-art approaches while incurring less complexity. The main contributions of this work and their novelties are listed as follows:
\begin{itemize}
    \item \textbf{Fully-Connected Hybrid Beamforming for AoD Estimation:} We consider the problem of fully-connected hybrid beamformer design with arbitrary array geometry and limited-resolution phase shifters, and develop an algorithm with lower complexity and superior performance compared to the best known state of the arts \cite{Lyu2021, Chen2017Aug, Ayach2014, Yu2016}. Additionally, using our proposed method as the initial point for the manifold optimization demonstrates performance improvements over both random initialization and the initial points derived from other state-of-the-art methods \cite{Lyu2021}. The novelty lies in the fact that such a problem has not yet been considered in the literature.
    \item \textbf{Theoretical Analysis of Quantization Error under Limited Resolution:} We provide theoretical analysis on the quantization error bound of the limited-resolution phase shifters, which is not available in existing literature of hybrid beamforming with limited-resolution phase shifters, e.g., \cite{Zhang2005, Ayach2014, Wang2018July, Lyu2021, Dong2019, Rajamaki2020, Zhang2014, Yu2016, Gao2016, Sohrabi2015, Sohrabi2016, Chen2017Aug, Lin2017, Wang2018May, Uwaechia2019, Li2020, Cheng2023, Cheng2023Aug}.
    \item \textbf{Convergence Analysis of the Proposed Algorithm:} We provide rigorous convergence analyses for the proposed algorithm. The novelties include: \textit{(i)} Our analyses differ from the related works in \cite{Hong2016, Boyd2011, Huang2016, Huang2023, Huang2023July, Wu2023} since our algorithm involves a quantization operation, which is not the case in the related works. \textit{(ii)} The convergence analyses presented in this paper go beyond our previous works in \cite{Huang2023, Huang2023July, Wu2023}, as the former additionally reveal that the point sequence produced by the proposed algorithm is a Cauchy sequence that converges to a fixed point after a finite number of iterations.
\end{itemize}

The remainder of this paper is organized as follows. The system model is described in Section \ref{SystemModel}. Section \ref{ProposedMethod} presents the proposed method for hybrid precoder design for AoD estimation. Section \ref{Section_analysis_ErrorBound_convergence} analyzes the quantization error bound and convergence behavior of the proposed algorithm. Various numerical examples are provided in Section \ref{NumericalResults} to demonstrate the effectiveness of the proposed approach, followed by conclusions in Section \ref{Conclusion}.

\textit{Notation:} We use italic letter (e.g., $N$) to denote scalar, bold-faced lower-case letter (e.g., ${\bf a}$) to denote column vector, bold-faced upper-case letter (e.g., ${\bf F}$) to denote matrix, and calligraphic or blackboard-bold letter (e.g., ${\mathcal{X}}$ and ${\mathbb C}$) to denote set. Further, ${\mathbb R}$ and ${\mathbb C}$ are the sets of real numbers and the set of complex numbers, respectively. $\jmath = \sqrt{-1}$ and $\Re\{\cdot\}$ returns the real part of a complex number. The superscripts $\cdot^{\textrm{T}}$, $\cdot^{\textrm{H}}$, $\cdot^{*}$, $\cdot^{-1}$, and $\cdot^{\dagger}$ denote transpose, Hermitian transpose, complex conjugate, inverse, and pseudo-inverse, respectively. $\textrm{diag}\{{\bf s}\}$ is the diagonal matrix with main diagonal being ${\bf s}$. $\mathcal{CN}({\bm \mu} , {\bf \Sigma})$ denotes the complex Gaussian distribution with mean ${\bm \mu}$ and variance ${\bf \Sigma}$. ${\bf I}$, ${\bf O}$, and ${\bf 1}$ are the identity matrix, all-zeros matrix, and all-ones vector, of appropriate sizes, respectively. $\|\cdot\|_{\text{F}}$ is the Frobenius norm of a matrix. For a given variable ${\bf F}$, we use the following notation for variables that relate to ${\bf F}$. We let ${\bf{\tilde F}}$, ${\bf{\widehat F}}$, and ${\bf F}^{\star}$ denote its auxiliary variable, an estimate variable by some algorithm, and the ideal variable (for example with infinite-resolution phase shifter). $\odot$ denotes the element-wise product. $[{\bf x}]_{i}$ and $[{\bf J}]_{ij}$ denote the $i$-th entry of vector ${\bf x}$ and the entry of matrix ${\bf J}$ at the $i$-th row and the $j$-th column. For a matrix ${\bf Z}$, $\text{rank({\bf Z})}$, $\text{tr}({\bf Z})$, and ${\bf Z} \succeq {\bf 0}$ denote the rank of ${\bf Z}$, the trace of ${\bf Z}$, and that ${\bf Z}$ is positive semidefinite.

\section{System Model}
\label{SystemModel}

\begin{figure}[t]
	\vspace*{-2mm}
	\centerline{\includegraphics[width=0.5\textwidth]{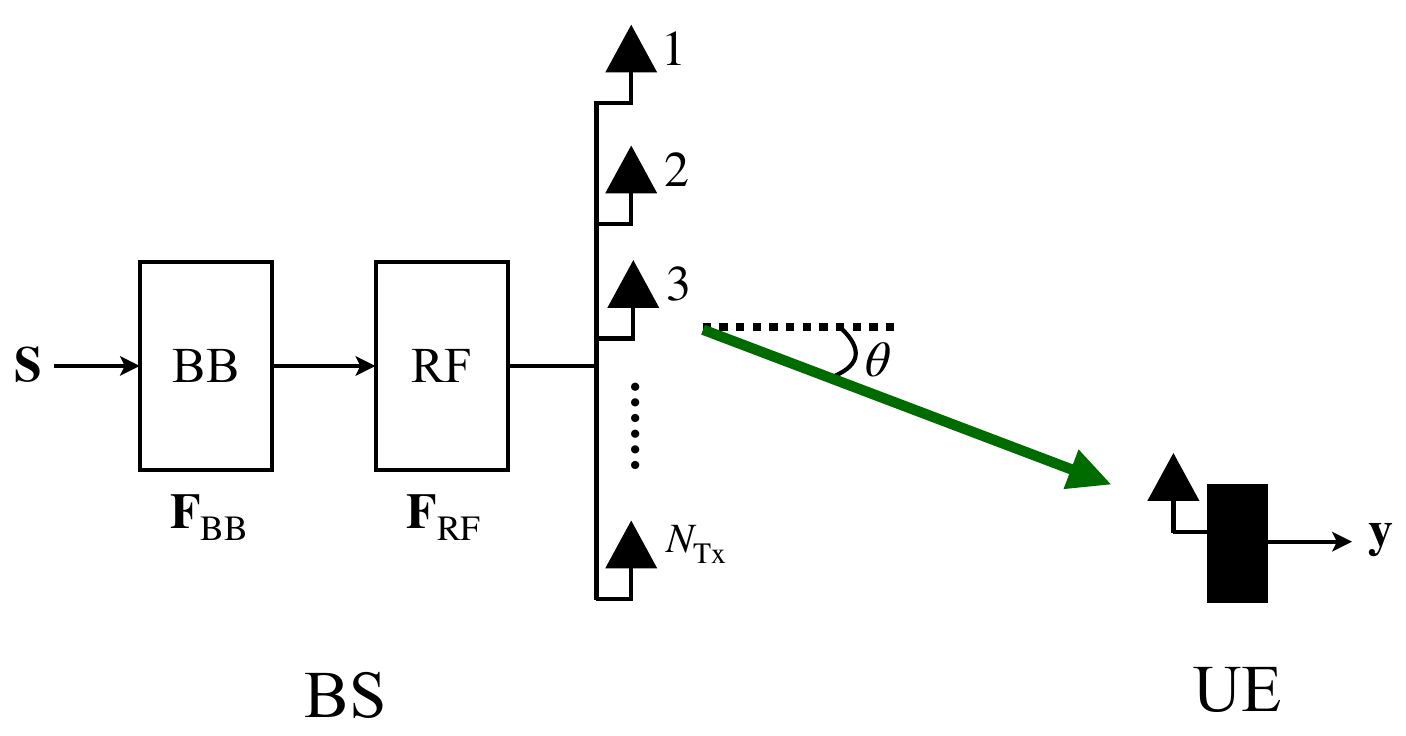}}
	\caption{Illustration of a mmWave downlink positioning scenario, where the single-antenna UE receives the pilot signals transmitted from the BS (consisting of a BB precoder, an RF precoder, and an arbitrary array of multiple antennas).}
	\label{system_model}
\end{figure}

We consider a mmWave downlink positioning scenario as in \cite{Fascista2019}, shown in Fig.~\ref{system_model}, where the base station (BS) consists of a BB precoder, an RF precoder, and an arbitrary array geometry of $N_{\text{Tx}}$ antennas; while the user equipment (UE) consists of a single antenna. The RF precoder is implemented by limited-resolution phase shifters, and the fully-connected structure is adopted, where each RF chain is connected to all antennas. Instead of partially-connected and arbitrarily-connected hybrid structures \cite{Lin2021, Wang2022Oct, Zhao2021, Hui2023, Liyanaarachchi2024}, we consider fully-connected hybrid structure as it can achieve the full beamforming gain for each RF chain \cite{Zhao2021}.

The BS transmits $M$ pilot symbols sequentially with identical power, denoted as $s_{m}, m = 1, 2, \cdots, M$. Employing a two-timescale hybrid precoding approach \cite{Hu2022, Liu2018June, Alkhateeb2014}, we adopt a transmission model in which the analog RF precoder is optimized at a slower time scale compared to the digital BB precoder. This prevents high hardware costs (attributed to rapid adaptation of the analog precoder) and reduces computational complexity, along with minimizing signaling overhead \cite{Hu2022}. In particular, each symbol is first precoded by a dedicated BB precoder vector, ${\bf f}_{\text{BB}, m} \in \mathbb{C}^{N_{\text{RF}}}$, and then precoded by an RF precoder constant for all symbols, ${\bf F}_{\text{RF}} \in \mathbb{C}^{N_{\text{Tx}} \times N_{\text{RF}}}$, where $N_{\text{RF}} \leq N_{\text{Tx}}$ denotes the number of RF chains. Considering highly directional mmWave transmissions, we assume a line-of-sight (LOS)-only channel.\footnote{The LOS path is resolvable from the non-line-of-sight paths due to channel sparsity, large number of antennas, and large bandwidth in the mmWave/THz wireless communication systems \cite{Chen2022}.} Thus, corresponding to the transmitted signal $s_{m}$, the received signal at the single-antenna UE can be modeled as
\begin{align}
\label{signal_model_y}
    y_{m} = \beta {\bf a}^{\textrm{T}}(\theta){\bf F}_{\text{RF}}{\bf f}_{\text{BB}, m}s_{m} + n_{m}, ~ m = 1, 2, \cdots, M,
\end{align}
where $\beta \in \mathbb{C}$ represents the complex channel gain, $\theta$ is the AoD, $n_{m}$ is the complex additive white Gaussian noise with zero mean and variance $\sigma_{\text{n}}^{2}$, and ${\bf a}(\theta)$ denotes the steering vector dependent on $\theta$ and the array geometry.

The signal model (\ref{signal_model_y}) can be written in vector form as\footnote{In our model, the number of pilots $M$ is independent from $N_{\text{Tx}}$ and $N_{\text{RF}}$, as can be seen in Sections \ref{ProposedMethod} and \ref{NumericalResults}. Hence, AoD can be estimated even when $M < N_{\text{RF}}$.}
\begin{align}
\label{signal_model_Y}
    {\bf y} = \beta {\bf S}({\bf F}_{\text{RF}}{\bf F}_{\text{BB}})^{\textrm{T}}{\bf a}(\theta) + {\bf n},
\end{align}
where ${\bf y} \! = \! [{y}_{1} , {y}_{2}, \cdots, {y}_{M}]^{\textrm{T}} \in \mathbb{C}^{M}$, ${\bf n} \! = \! [{n}_{1}, {n}_{2}, \cdots, {n}_{M}]^{\textrm{T}} \in \mathbb{C}^{M}$, ${\bf F}_{\text{BB}} = [{\bf f}_{\text{BB}, 1}, {\bf f}_{\text{BB}, 2}, \cdots, {\bf f}_{\text{BB}, M}] \in \mathbb{C}^{N_{\text{RF}} \times M}$, and ${\bf S} = \textrm{diag}\{s_{1}, s_{2}, \cdots, s_{M}\}$. In addition, ${\bf n} \sim \mathcal{CN}({\bf 0} , \sigma_{\text{n}}^{2}{\bf I})$ with known noise power $\sigma_{\text{n}}^{2}$. Our goal is to design an RF precoder and a BB precoder such that the accuracy of estimation of AoD is maximized, under the BS transmit power constraint and the hardware constraint of limited-resolution phase shifters.

\begin{remark}
    We consider the case of single UE equipped with single-antenna without loss of generality.  The framework can be easily extended to a UE with multiple antennas, i.e.,
    \begin{align}
    \label{signal_model_Y_multUE}
    {\bf Y} = \beta {\bf S}({\bf F}_{\textnormal{RF}}{\bf F}_{\textnormal{BB}})^{{\textnormal{\textrm{T}}}}{\bf a}(\theta) {\bf b}^{{\textnormal{\textrm{T}}}}(\phi) + {\bf N},
\end{align}
where ${\bf Y}$ is the observation matrix where each row corresponds to the observations at the UE array of a single transmission and ${\bf b}(\phi)$ denotes the steering vector at the UE as a function of angle-of-arrival $\phi$.
    It can be seen from \eqref{signal_model_Y_multUE} that the difference compared to \eqref{signal_model_Y} is that a multiple-antenna UE can provide a gain in the signal-to-noise ratio (SNR), while the fundamental principle of AoD estimation remains the same. 
    In addition, our framework can be extended to the case of multiple UEs, which is further elaborated in Remark \ref{remark_multipleUE}, after we have introduced the codebook design in Section \ref{ProposedTwoStepStrategy}.
\end{remark}

\section{Proposed Method}
\label{ProposedMethod}

\subsection{CRB-Based Performance Metric}
\label{Proposed_CRB}
Define ${\bf{\tilde{y}}} \triangleq \beta {\bf S}({\bf F}_{\text{RF}}{\bf F}_{\text{BB}})^{\textrm{T}}{\bf a}(\theta)$. Then, the Fisher information matrix  ${\bf J}({\bf F}_{\text{RF}}, {\bf F}_{\text{BB}}; {\bf x}) \in \mathbb{R}^{3 \times 3}$ can be computed by using the Slepian-Bangs formula \cite{Kay2017} as
\begin{align}
\label{FIM}
    [{\bf J}]_{ij} = \frac{2}{\sigma_{\text{n}}^{2}} \Re \left\{ \left( \frac{\partial {\bf{\tilde{y}}} }{\partial [{\bf x}]_{i}} \right)^{\textrm{H}} \left( \frac{\partial {\bf{\tilde{y}}} }{\partial [{\bf x}]_{j}} \right) \right\},
\end{align}
where ${\bf x} = [\theta, \beta_{\text{R}}, \beta_{\text{I}}]^{\textrm{T}}$ contains all the unknown parameters. In addition, $\beta_{\text{R}}$ and $\beta_{\text{I}}$ denote the real and imaginary parts of $\beta$, respectively. The derivative of ${\bf{\tilde{y}}}$ with respect to (w.r.t.) $[{\bf x}]_{i}$ is calculated as in Appendix \ref{appendix_A}. The corresponding CRB matrix is defined as
\begin{align}
\label{CRB}
    {\bf C} = {\bf J}^{-1}.
\end{align}

To quantify the AoD estimation accuracy, we adopt the angle error bound (AEB) as our performance metric, computed as \eqref{AEB} displayed at the top of this page, where $\sigma_{\text{s}}$ is the signal power, ${\bf D} \triangleq \text{diag}\{0, 1, \cdots, N_{\text{Tx}-1}\}$, ${\bf F} = {\bf F}_{\text{RF}} {\bf F}_{\text{BB}}$, and we have employed the block matrix inversion lemma \cite{Anton1994} as detailed in Appendix \ref{appendix_B}. 

\begin{figure*}[t]
\vspace{-0.3cm}
\begin{align} \tag{6}
\label{AEB}
    \text{AEB}({\bf F}_{\text{RF}}, {\bf F}_{\text{BB}}; {\bf x})
    = \sqrt{[{\bf C}]_{11}} = \frac{\sigma_{\text{n}}}{\sigma_{\text{s}}} \frac{\lambda}{2\sqrt{2} |\beta| \pi d} \sqrt{ \frac{{\bf a}^{\textrm{H}}(\theta){\bf F}^{*}{\bf F}^{\textrm{T}}{\bf a}(\theta) }{
        {\bf a}^{\textrm{H}}(\theta){\bf F}^{*}{\bf F}^{\textrm{T}} \big[ {\bf a}(\theta){\bf a}^{\textrm{H}}(\theta){\bf D}{\bf F}^{*}{\bf F}^{\textrm{T}}{\bf D} - {\bf D}{\bf a}(\theta){\bf a}^{\textrm{H}}(\theta){\bf D}{\bf F}^{*}{\bf F}^{\textrm{T}} \big]{\bf a}(\theta) } }
\end{align}
\end{figure*}

\setcounter{equation}{6}

\subsection{Problem Formulation for Optimal Precoder Design}
\label{Section_ProblemFormulationOpt}
The AEB depends on the unknown parameters in ${\bf x}$. We assume that ${\bf x}$ belongs to an uncertainty set $\mathcal{X}$ that can be, e.g., determined via some tracking algorithms \cite{Garcia2016, Garcia2018, Keskin2022}. For any ${\bf x} \in \mathcal{X}$, the AEB is only a function of ${\bf F}_{\text{RF}}$ and ${\bf F}_{\text{BB}}$. The optimal precoder design problem can be formulated as
\begin{subequations}
\label{OptPrecoder_problem}
\begin{align}
\label{objective_OptPrecoder_problem}
    \min_{{\bf F}_{\text{RF}}, {\bf F}_{\text{BB}}} & ~ \text{AEB}({\bf F}_{\text{RF}}, {\bf F}_{\text{BB}}; {\bf x}) \\
\label{constraint1_OptPrecoder_problem}
    \text{s.t.} ~~\! & ~ \|{\bf F}_{\text{RF}}{\bf F}_{\text{BB}}\|_{\text{F}}^{2} = P,  \\
\label{constraint2_OptPrecoder_problem}
    & ~ \left[{\bf F}_{\text{RF}}\right]_{ij} \in \mathcal{F}, ~~ 1 \leq i \leq N_{\text{Tx}} ~\text{and}~ 1 \leq j \leq N_{\text{RF}},
\end{align}
\end{subequations}
where $P$ stands for the total transmit power of the BS antennas, and $\mathcal{F}$ denotes the set for limited resolution of the phase shifters, which is defined as:
\begin{align}
\label{PS_feasibleset}
    \mathcal{F} \triangleq \left\{ \frac{1}{\sqrt{N_{\text{Tx}}}} e^{\jmath 2 \pi b / 2^{B}} \bigg| b = 0, 1, \cdots, 2^{B} \!-\! 1 \right\},
\end{align}
with $B$ representing the total number of quantization bits of the phase shifters.

\subsection{Two-Step Strategy for Solving Problem~\eqref{OptPrecoder_problem}}
\label{ProposedTwoStepStrategy}
It is difficult to directly solve Problem \eqref{OptPrecoder_problem} w.r.t. ${\bf F}_{\text{RF}}$ and ${\bf F}_{\text{BB}}$, due to the complicated structure\footnote{The denominator of $\text{AEB}({\bf F}_{\text{RF}}, {\bf F}_{\text{BB}}; {\bf x})$ contains quartic terms w.r.t. ${\bf F} = {\bf F}_{\text{RF}} {\bf F}_{\text{BB}}$.} of $\text{AEB}({\bf F}_{\text{RF}}, {\bf F}_{\text{BB}}; {\bf x})$ and the discrete-phase nature of the entries of ${\bf F}_{\text{RF}}$. We provide a strategy for solving Problem (\ref{OptPrecoder_problem}) via the following two steps:
\begin{itemize}
    \item \textbf{Step~1:} Finding the approximately optimal fully digital precoder ${\bf F}_{\text{opt}}$ as a solution to Problem (\ref{OptPrecoder_problem}).
    \item \textbf{Step~2:} Finding a decomposition of ${\bf F}_{\text{opt}}$ to obtain the best approximation ${\bf F}_{\text{opt}}\approx {\bf F}_{\text{RF}}{\bf F}_{\text{BB}}$ in the least-squares (LS) sense.
\end{itemize}
We now elaborate on these two steps.

\textbf{Step 1:} Based on the fact that the unknown variables ${\bf F}_{\text{RF}}$ and ${\bf F}_{\text{BB}}$ appear as a product (i.e., ${\bf F}_{\text{RF}}{\bf F}_{\text{BB}}$) in both the objective function (\ref{objective_OptPrecoder_problem}) and the constraint (\ref{constraint1_OptPrecoder_problem}), for any ${\bf x} \in \mathcal{X}$, we consider the following optimization problem:
\begin{align}
\label{OptPrecoder_problem_F}
    \min_{{\bf F}} ~ \text{AEB}({\bf F}; {\bf x}) \quad 
    \text{s.t.} ~ \|{\bf F}\|_{\text{F}}^{2} = P,
\end{align}
where ${\bf F} = {\bf F}_{\text{RF}}{\bf F}_{\text{BB}} \in \mathbb{C}^{N_{\text{Tx}} \times M}$ and we drop the constraint (\ref{constraint2_OptPrecoder_problem}) temporarily. This corresponds to a fully digital precoder optimization \cite{Keskin2022}. We define ${\bf Z} \triangleq {\bf F}{\bf F}^{\textrm{H}}$, and equivalently reformulate Problem \eqref{OptPrecoder_problem_F} as:
\begingroup
\allowdisplaybreaks
\begin{subequations}
\begin{align}
    \min_{{\bf Z}, u} ~\! u ~~ \text{s.t.} & \left[ \begin{array}{cc}
    {\bf J}({\bf Z}; {\bf x}) & {\bf e}_{1} \\
    {\bf e}_{1}^{\textrm{T}}  & u
    \end{array}\right] \succeq {\bf 0}, \\
    & ~\! \text{tr}({\bf Z}) = P, ~ {\bf Z} \succeq {\bf 0}, ~ \text{rank}({\bf Z}) = M,
\end{align}
\end{subequations}
\endgroup
where ${\bf e}_{1} = [1, 0, 0]^{\text{T}}$. Taking into account the uncertainty of ${\bf x}$, i.e., ${\bf x} \in \mathcal{X}$, and by discretizing ${\mathcal{X}}$ into a uniform grid of $G$ points $\{{\bf x}_{g}\}_{g = 1}^{G}$, a robust version of the above problem can be expressed as
\begingroup
\allowdisplaybreaks
\begin{subequations}
\label{Robust_OptPrecoder_problem}
\begin{align}
    \min_{{\bf Z}, \{u_{g}\}} & ~\! \max_{{\bf x}_{g} \in \mathcal{X}} ~\! u_{g} \\
    \text{s.t.} ~~ & \left[ \begin{array}{cc}
    {\bf J}({\bf Z}; {\bf x}_{g}) & {\bf e}_{1} \\
    {\bf e}_{1}^{\textrm{T}}  & u_{g}
    \end{array} \right] \succeq {\bf 0}, ~\! g = 1, 2, \cdots, G, \\
    & ~\! \text{tr}({\bf Z}) = P, ~ {\bf Z} \succeq {\bf 0}, ~ \text{rank}({\bf Z}) = M.
\end{align}
\end{subequations}
\endgroup
Problem \eqref{Robust_OptPrecoder_problem} can equivalently be reformulated as
\begingroup
\allowdisplaybreaks
\begin{subequations}
\label{Discreze_OptPrecoder_problem_F}
\begin{align}
    \min_{{\bf Z}, \{u_{g}\}, t} & ~\! t \\
    \text{s.t.} ~~ & \left[ \begin{array}{cc}
    {\bf J}({\bf Z}; {\bf x}_{g}) & {\bf e}_{1} \\
    {\bf e}_{1}^{\textrm{T}}  & u_{g}
    \end{array} \right] \succeq {\bf 0}, ~\! g = 1, 2, \cdots, G, \\
    & ~\! u_{g} \leq t, ~\! g = 1, 2, \cdots, G, \\
    & ~\! \text{tr}({\bf Z}) = P, ~ {\bf Z} \succeq {\bf 0}, ~ \text{rank}({\bf Z}) = M.
\end{align}
\end{subequations}
\endgroup
It is shown in \cite{Keskin2022, Nickel2006, Li2008} that a codebook-based approach can be applied to decrease the complexity while achieving a satisfactory solution. Specifically, a predefined codebook consists of directional and derivative beams \cite{Keskin2022}, that is, ${\bf F}^{(\text{pre})} = [{\bf F}^{(\text{direc})} , {\bf F}^{(\text{deriv})}]$, where ${\bf F}^{(\text{direc})} = [{\bf a}(\theta_{1}), {\bf a}(\theta_{2}), \cdots, {\bf a}(\theta_{G})]$ and ${\bf F}^{(\text{deriv})} = [{\bf{\dot a}}(\theta_{1}), {\bf{\dot{a}}}(\theta_{2}), \cdots, {\bf{\dot{a}}}(\theta_{G})]$, with $\theta_{g}$ corresponding to ${\bf x}_{g}$ for $g = 1, 2, \cdots, G$, ${\bf{\dot a}}(\theta) = \frac{\partial {\bf a}(\theta)}{\partial \theta}$ and $G = M/2$. With the predefined codebook ${\bf F}^{(\text{pre})}$, we consider the optimal beam power allocation problem in ${\bf q} = [q_{1}, q_{2}, \cdots, q_{M}]^{\textrm{T}}$:
\begingroup
\allowdisplaybreaks
\begin{subequations}
\label{Discreze_OptPrecoder_problem_F}
\begin{align}
    \min_{{\bf q}, \{u_{g}\}, t} & ~\! t \\
    \text{s.t.} ~~ & \left[ \begin{array}{cc}
    {\bf J}( {\bf F}^{(\text{pre})}\text{diag}\{{\bf q}\}({\bf F}^{(\text{pre})})^{\textrm{H}} ; {\bf x}_{g}) & {\bf e}_{1} \\
    {\bf e}_{1}^{\textrm{T}}  & u_{g}
    \end{array} \right] \succeq {\bf 0}, \\
    & ~\! u_{g} \leq t, ~~ g = 1, 2, \cdots, G, \\
    & ~\! \text{tr}({\bf F}^{(\text{pre})}\text{diag}\{{\bf q}\}({\bf F}^{(\text{pre})})^{\textrm{H}}) = P,  
\end{align}
\end{subequations}
\endgroup
where we have made use of ${\bf Z} = {\bf F}^{(\text{pre})}\text{diag}\{{\bf q}\}({\bf F}^{(\text{pre})})^{\textrm{H}}$ and the fact that both ${\bf F}^{(\text{pre})}\text{diag}\{{\bf q}\}({\bf F}^{(\text{pre})})^{\textrm{H}} \succeq {\bf 0}$ and $\text{rank}({\bf F}^{(\text{pre})}\text{diag}\{{\bf q}\}({\bf F}^{(\text{pre})})^{\textrm{H}}) = M$ are inherently satisfied (and thus omitted from \eqref{Discreze_OptPrecoder_problem_F}).
Finally, \eqref{Discreze_OptPrecoder_problem_F} yields the optimal fully digital precoder as
\begin{align}
\label{OptimalFDPrecoder}
    {\bf F}_{\text{opt}} = {\bf F}^{(\text{pre})}{\text{diag}}\{\sqrt{q_{1}}, \sqrt{q_{2}}, \cdots, \sqrt{q}_{M}\}.
\end{align}

\begin{remark}
\label{remark_multipleUE}
    The above codebook-based approach can be extended to the case of multiple UEs (say $L$ UEs). To this end, we assign ${\bf F}^{(\textnormal{pre})} = [{\bf F}^{(\textnormal{direc})}_{(1)} , {\bf F}^{(\textnormal{deriv})}_{(1)} , \cdots, {\bf F}^{(\textnormal{direc})}_{(L)} , {\bf F}^{(\textnormal{deriv})}_{(L)}]$, where ${\bf F}^{(\textnormal{direc})}_{(l)}$ and ${\bf F}^{(\textnormal{deriv})}_{(l)}$ are the directional and derivative beams corresponding to the $l$-th UE, for $l = 1, 2, \cdots, L$. In this case, the number of columns of ${\bf F}^{(\textnormal{direc})}_{(l)}$ and ${\bf F}^{(\textnormal{deriv})}_{(l)}$ equals to $M / (2L)$. With this ${\bf F}^{(\textnormal{pre})}$, we solve Problem \eqref{Discreze_OptPrecoder_problem_F} and obtain the optimal fully digital precoder for $L$ UEs via \eqref{OptimalFDPrecoder}. Further, note that the number of columns of ${\bf F}^{(\textnormal{direc})}_{(l)}$ and ${\bf F}^{(\textnormal{deriv})}_{(l)}$ should be greater than or equal to 1. We have $M/(2L) \geq 1$, that is $M \geq 2L$. In other words, the number of pilots should be greater than or equal to 2 times the number of UEs.  
\end{remark}

\textbf{Step 2:} We decompose ${\bf F}_{\text{opt}}$ into two matrices, i.e., ${\bf F}_{\text{RF}}$ and ${\bf F}_{\text{BB}}$, by taking into account the constraints \eqref{constraint1_OptPrecoder_problem} and (\ref{constraint2_OptPrecoder_problem}):
\begin{align}
\label{OptPrecoder_problem_FF}
    \min_{{\bf F}_{\text{RF}}, {\bf F}_{\text{BB}}} ~ \frac{1}{2}\|{\bf F}_{\text{opt}} - {\bf F}_{\text{RF}}{\bf F}_{\text{BB}}\|_{\text{F}}^{2} \quad 
    \text{s.t.} ~ \text{(\ref{constraint1_OptPrecoder_problem})} ~ \text{and} ~ \text{(\ref{constraint2_OptPrecoder_problem})}.
\end{align}
In what follows, we propose an alternating optimization approach for solving Problem (\ref{OptPrecoder_problem_FF}). To be specific, we first solve ${\bf F}_{\text{BB}}$ with a fixed ${\bf F}_{\text{RF}}$, as
\begin{align}
\label{OptPrecoder_problem_BB}
    \min_{{\bf F}_{\text{BB}}} ~ \frac{1}{2}\|{\bf F}_{\text{opt}} - {\bf F}_{\text{RF}}{\bf F}_{\text{BB}}\|_{\text{F}}^{2} \quad
    \text{s.t.} ~ \text{(\ref{constraint1_OptPrecoder_problem})}.
\end{align}
It has a LS closed-form solution as
\begin{align}
\label{solution_F_BB}
    {\bf F}_{\text{BB}} = \frac{\sqrt{P}}{\|{\bf F}_{\text{RF}}{\bf F}_{\text{RF}}^{\dagger}{\bf F}_{\text{opt}}\|_{\text{F}}} {\bf F}_{\text{RF}}^{\dagger}{\bf F}_{\text{opt}},
\end{align}
where ${\bf F}_{\text{RF}}^{\dagger} = ({\bf F}_{\text{RF}}^{\textrm{H}}{\bf F}_{\text{RF}})^{-1}{\bf F}_{\text{RF}}^{\textrm{H}}$. Then, we solve ${\bf F}_{\text{RF}}$ with the obtained ${\bf F}_{\text{BB}}$ in (\ref{solution_F_BB}), as
\begin{align}
\label{OptPrecoder_problem_RF}
    \min_{{\bf F}_{\text{RF}}} ~ \frac{1}{2}\|{\bf F}_{\text{opt}} - {\bf F}_{\text{RF}}{\bf F}_{\text{BB}}\|_{\text{F}}^{2} \quad
    \text{s.t.} ~ \text{(\ref{constraint2_OptPrecoder_problem})}.
\end{align}
We develop an algorithm based on the ADMM \cite{Boyd2011} to solve the above problem. To this end, we introduce an auxiliary variable ${\bf{\tilde F}}_{\text{RF}} \in \mathbb{C}^{N_{\text{Tx}} \times N_{\text{RF}}}$, and Problem (\ref{OptPrecoder_problem_RF}) can be equivalently expressed as
\begin{align}
\label{OptPrecoder_problem_RF_auxiliary}
    \min_{{\bf F}_{\text{RF}}, {\bf{\tilde F}}_{\text{RF}}} ~\!\! \frac{1}{2}\|{\bf F}_{\text{opt}} - {\bf{\tilde F}}_{\text{RF}}{\bf F}_{\text{BB}}\|_{\text{F}}^{2} \quad
    \text{s.t.} ~ \text{(\ref{constraint2_OptPrecoder_problem})} ~\text{and}~ {\bf{\tilde F}}_{\text{RF}} = {\bf F}_{\text{RF}}.
\end{align}
The corresponding scaled-form augmented Lagrangian function is given as \cite{Boyd2011}
\begin{align}
\label{AugLagFun}
    \mathcal{L}\Big({\bf{\tilde F}}_{\text{RF}}, {\bf F}_{\text{RF}}, & ~\! {\bf U}\Big) = \frac{1}{2}\|{\bf F}_{\text{opt}} - {\bf{\tilde F}}_{\text{RF}}{\bf F}_{\text{BB}}\|_{\text{F}}^{2} \nonumber \\ 
    & + \frac{\rho}{2}\left( \|{\bf{\tilde F}}_{\text{RF}} - {\bf F}_{\text{RF}} + {\bf U}\|_{\text{F}}^{2} - \|{\bf U}\|_{\text{F}}^{2} \right),
\end{align}
where ${\bf U} \in \mathbb{C}^{N_{\text{Tx}} \times N_{\text{RF}}}$ is the scaled dual variable and $\rho > 0$ is the augmented Lagrangian parameter. Parameter $\rho$ can be set based on the proposed convergence analyses in Section \ref{ConvergenceAnalysis}. The primal, auxiliary, and dual variables are updated as:
\begin{subequations}
\label{Update_F_U}
    \begin{align}
        \label{Update_F_RF}
        {\bf F}_{\text{RF}}^{(\! k \!+\! 1 \!)} & = \argmin_{\left[{\bf F}_{\text{RF}}\right]_{ij} \in \mathcal{F}} ~ \mathcal{L}({\bf{\tilde F}}_{\text{RF}}^{(\! k \!)}, {\bf F}_{\text{RF}}, {\bf U}^{(\! k \!)})  \nonumber \\
        & = \frac{1}{\sqrt{N_{\text{Tx}}}} e^{\jmath \mathcal{Q}\left( \angle({\bf{\tilde F}}_{\text{RF}}^{(\! k \!)} + {\bf U}^{(\! k \!)}) \right) },   \\
        \label{Update_F_RF_tilde}
        {\bf{\tilde F}}_{\text{RF}}^{(\! k \!+\! 1 \!)} & = \argmin_{{\bf{\tilde F}}_{\text{RF}}} ~ \mathcal{L}({\bf{\tilde F}}_{\text{RF}}, {\bf F}_{\text{RF}}^{(\! k \!+\! 1 \!)}, {\bf U}^{(\! k \!)}) \nonumber \\
        & = [{\bf F}_{\text{opt}}{\bf F}_{\text{BB}}^{\textrm{H}} \!+\! \rho ({\bf F}_{\text{RF}}^{(\! k \!+\! 1 \!)} \!\!-\! {\bf U}^{(\! k \!)} \! )]({\bf F}_{\text{BB}}{\bf F}_{\text{BB}}^{\textrm{H}} \!+\! \rho{\bf I})^{-1},  \\
        \label{Update_U}
        {\bf U}^{(\! k \!+\! 1 \!)} & = {\bf U}^{(\! k \!)} + {\bf{\tilde F}}_{\text{RF}}^{(\! k \!+\! 1 \!)} - {\bf F}_{\text{RF}}^{(\! k \!+\! 1 \!)}.
    \end{align}
\end{subequations}
In (\ref{Update_F_RF}), $\angle \cdot$ denotes the angle of its argument in an element-wise manner, and ${\mathcal{Q}}(\cdot)$ stands for the quantization function rounding its argument to the available phases of the phase shifters $\bigl($i.e., $\frac{2\pi}{2^{B}} \! \times \! \left\{ 0, 1, \cdots, 2^{B} \!-\! 1 \right\}$$\bigr)$. 

The proposed algorithm for solving Problem (\ref{OptPrecoder_problem_FF}) is referred to as AltOpt-LS-ADMM, and summarized in Algorithm \ref{Proposed_Alg}, where superscript $\cdot^{(\!i\!)}$ denotes the corresponding variable at the $i$-th outer iteration, superscript $\cdot^{(\!k\!)}$ denotes the corresponding variable at the $k$-th inner (i.e., ADMM) iteration, and $I_{\text{max}}$ and $k_{\text{max}}$ are the maximal numbers of the outer and the inner loops, respectively. Besides, ${\bf F}_{\text{RF}}^{(\text{init})}$ and ${\bf{\tilde F}}_{\text{RF}}^{(\text{init})}$ are obtained by randomly selecting from the feasible set \eqref{PS_feasibleset}, the update of $\rho$ in Line 3 comes from \eqref{rho_convergent_condition} in Section \ref{ConvergenceAnalysis}, and ${\bf O}$ in Line 4 is an all-zeros matrix.

\begin{algorithm}[t]
	\caption{AltOpt-LS-ADMM for solving Problem (\ref{OptPrecoder_problem_FF})}
	\label{Proposed_Alg}
	\textbf{Input~~~~\!:} ${\bf F}_{\text{opt}} \in \mathbb{C}^{N_{\text{Tx}} \times M}$, $I_{\text{max}}$, $k_{\text{max}}$ \\
	\textbf{Output~~\!:} ${\bf F}_{\text{RF}} \in \mathbb{C}^{N_{\text{Tx}} \times N_{\text{RF}}}$, ${\bf F}_{\text{BB}} \in \mathbb{C}^{N_{\text{RF}} \times M}$ \\
	\textbf{Initialize:} ${\bf F}_{\text{RF}}^{(\!0\!)} = {\bf F}_{\text{RF}}^{(\text{init})}$, $i = 0$
	\begin{algorithmic}[1]
		\While {$i < I_{\text{max}}$} \vspace{1mm} \\
		 ~~~ ${\bf F}_{\text{BB}}^{(\!i \!+\! 1\!)} = \frac{\sqrt{P}}{\|{\bf F}_{\text{RF}}^{(\!i\!)}{\bf F}_{\text{RF}}^{(\!i\!)\dagger}{\bf F}_{\text{opt}}\|_{\text{F}}} {\bf F}_{\text{RF}}^{(\!i\!)\dagger}{\bf F}_{\text{opt}}$  \vspace{1mm}  \\
        ~~~ $\rho = \mathrm{max} \left\{ \sqrt{2}\|{\bf F}_{\text{{BB}}}^{(\!i + 1\!)}({\bf F}_{\text{{BB}}}^{(\!i + 1\!)})^{\textrm{{H}}}\|_{\text{{F}}} ~\! , \|{\bf F}_{\text{{BB}}}^{(\!i + 1\!)}\|_{\text{{F}}}^{2} \right\}$  \vspace{1mm} \\
        ~~~ ${\bf{\tilde F}}_{\text{RF}}^{(\!0\!)} \!=\! {\bf{\tilde F}}_{\text{RF}}^{(\text{init})}$\!, \! ${\bf U}^{(\!0\!)} \!=\! {\bf O}$, \! $k \!=\! 0$ \Comment{ADMM initialization} \vspace{1mm}
        \Repeat  \vspace{1mm} \\
        ~~~~~~ ${\bf F}_{\text{RF}}^{(\!k + 1\!)} \gets $ \eqref{Update_F_RF}  \vspace{1mm} \\
        ~~~~~~ ${\bf{\tilde{F}}}_{\text{RF}}^{(\! k + 1 \!)} \gets $ \eqref{Update_F_RF_tilde}  \vspace{1mm}  \\
        ~~~~~~ ${\bf U}^{(\! k + 1 \!)} \gets $ \eqref{Update_U}  \vspace{1mm} \\
        ~~~~~~~\! $k = k + 1$
        \Until{$k = k_{\text{max}}$}  \vspace{1mm} \\
        ~~~ ${\bf F}_{\text{RF}}^{(\!i + 1 \!)} = \frac{1}{\sqrt{N}_{\text{Tx}}}e^{\jmath \mathcal{Q}(\angle {\bf{\tilde{F}}}_{\text{RF}}^{(\!k\!)})} $  \vspace{1mm} \\
		~~~ $i = i+1$
		\EndWhile   \vspace{1mm} \\
		${\bf F}_{\text{RF}} = {\bf F}_{\text{RF}}^{(\!i\!)}$ and ${\bf F}_{\text{BB}} = {\bf F}_{\text{BB}}^{(\!i\!)}$
	\end{algorithmic}
\end{algorithm}

\begin{remark}
The proposed two-step strategy can find approximate (but not exact) solutions to the original optimal precoder design problem, i.e., Problem (\ref{OptPrecoder_problem}).
\end{remark}

\subsection{Computational Complexity Analysis}
The computational cost of the proposed AltOpt-LS-ADMM algorithm mainly comes from the pseudo-inverse operation and the multiplication operation in Line 2 and the inverse operation and the multiplication operation in Line 7, which incur the complexities $\mathcal{O}(N_{\text{RF}}^{2}N_{\text{Tx}})$, $\mathcal{O}(N_{\text{RF}}N_{\text{Tx}}M)$, $\mathcal{O}(N_{\text{RF}}^{3})$, and $\mathcal{O}(N_{\text{RF}}N_{\text{Tx}}M)$, respectively. Since we can compute the inverse operation in Line 7 outside the ADMM iteration and then use its results for all inner iterations, the total computational cost of the proposed AltOpt-LS-ADMM algorithm is $\mathcal{O}(I_{\text{max}}(N_{\text{RF}}^{2}N_{\text{Tx}} + (1 + k_{\text{max}})N_{\text{RF}}N_{\text{Tx}}M))$.

\section{Analysis of Error Bounds and Convergence}
\label{Section_analysis_ErrorBound_convergence}
\subsection{Analysis of Quantization Error Bound}
\label{AnalysisofQuanErrorBound}
In this subsection, we analyse the quantization error bound in the proposed ADMM algorithm (i.e., inner iteration of Algorithm~\ref{Proposed_Alg}) resulting from the quantization operation in Line 6 of Algorithm \ref{Proposed_Alg}. We first denote ${\bf{\widehat F}}_{\text{RF}}$ and ${\bf F}_{\text{RF}}^{\star}$ as the RF precoder with (i.e., $B < \infty$) and without (i.e., $B = \infty$) quantization, respectively. Then, the relation of these two matrices is given as
${\bf{\widehat F}}_{\text{RF}} = {\bf \Phi} \odot {\bf F}_{\text{RF}}^{\star}$, where ${\bf \Phi} \in \mathbb{C}^{N_{\text{Tx}} \times N_{\text{RF}}}$ is the quantization error matrix. Moreover, the elements of ${\bf \Phi}$ can be formulated as $[{\bf \Phi}]_{ij} \!=\! e^{\jmath \phi_{ij}}$, where $0 \!\leq\! |\phi_{ij}| \!\leq\! \pi/2^{B}$ for all $1 \leq i \leq N_{\text{Tx}}$ and $1 \leq j \leq N_{\text{RF}}$. Therefore, the quantization error can be calculated as
\begin{subequations}
\label{QuantizationIn}
\begin{align}
    & \|{\bf F}_{\text{opt}} - {\bf{\widehat{F}}}_{\text{RF}}{\bf F}_{\text{BB}}\|_{\text{F}} - \|{\bf F}_{\text{opt}} - {\bf F}_{\text{RF}}^{\star}{\bf F}_{\text{BB}}\|_{\text{F}} \nonumber \\
    = & ~\! \|{\bf F}_{\text{opt}} - ({\bf \Phi} \odot {\bf F}_{\text{RF}}^{\star}) {\bf F}_{\text{BB}}\|_{\text{F}} - \|{\bf F}_{\text{opt}} - {\bf F}_{\text{RF}}^{\star}{\bf F}_{\text{BB}}\|_{\text{F}} \nonumber \\
    \label{QuantizationIn_b}
    \leq & ~\! \| ({\bf F}_{\text{RF}}^{\star} - ({\bf \Phi} \odot {\bf F}_{\text{RF}}^{\star})) {\bf F}_{\text{BB}} \|_{\text{F}} \\
    = & ~\! \|[({\bf 1} - {\bf \Phi}) \odot {\bf F}_{\text{RF}}^{\star}] {\bf F}_{\text{BB}}\|_{\text{F}} \nonumber \\
    \label{QuantizationIn_d}
    \leq & ~\! \|({\bf 1} - {\bf \Phi}) \odot {\bf F}_{\text{RF}}^{\star}\|_{\text{F}} \|{\bf F}_{\text{BB}}\|_{\text{F}}  \\
    \label{QuantizationIn_e}
    \leq & ~\! \|{\bf 1} - {\bf \Phi}\|_{\text{F}} \|{\bf F}_{\text{RF}}^{\star}\|_{\text{F}} \|{\bf F}_{\text{BB}}\|_{\text{F}}  \\
    \label{QuantizationIn_f}
    \leq & ~\! \left|1 - e^{\jmath {\pi}/{2^{B}}} \right| \sqrt{N_{\text{Tx}}N_{\text{RF}}} \|{\bf F}_{\text{RF}}^{\star}\|_{\text{F}} \|{\bf F}_{\text{BB}}\|_{\text{F}},
\end{align}
\end{subequations}
where in \eqref{QuantizationIn_b} we used the triangle inequality, in \eqref{QuantizationIn_d} we employed the fact that $\|{\bf M}{\bf N}\|_{\text{F}} \leq \|{\bf M}\|_{\text{F}}\|{\bf N}\|_{\text{F}}$ holds for any matrices ${\bf M}$ and ${\bf N}$ of appropriate sizes, in \eqref{QuantizationIn_e} we utilized the Cauchy-Schwarz inequality, and in \eqref{QuantizationIn_f} we used the following inequality:
\begin{align*}
    \|{\bf 1} - {\bf \Phi}\|_{\text{F}} = & ~\! \sqrt{ \sum_{i = 1}^{N_{\text{Tx}}} \sum_{j = 1}^{N_{\text{RF}}} \left| 1 - e^{\jmath \phi_{ij}} \right|^{2}  }  \\
    \leq & ~\! \sqrt{ \sum_{i = 1}^{N_{\text{Tx}}} \sum_{j = 1}^{N_{\text{RF}}} \left| 1 - e^{\jmath \pi/2^{B}} \right|^{2}  } \\
    = & ~\! \left|1 - e^{\jmath {\pi}/{2^{B}}} \right| \sqrt{N_{\text{Tx}}N_{\text{RF}}} ~\!.
\end{align*}

\begin{figure}[t]
	\vspace*{-2mm}
	\centerline{\includegraphics[width=0.5\textwidth]{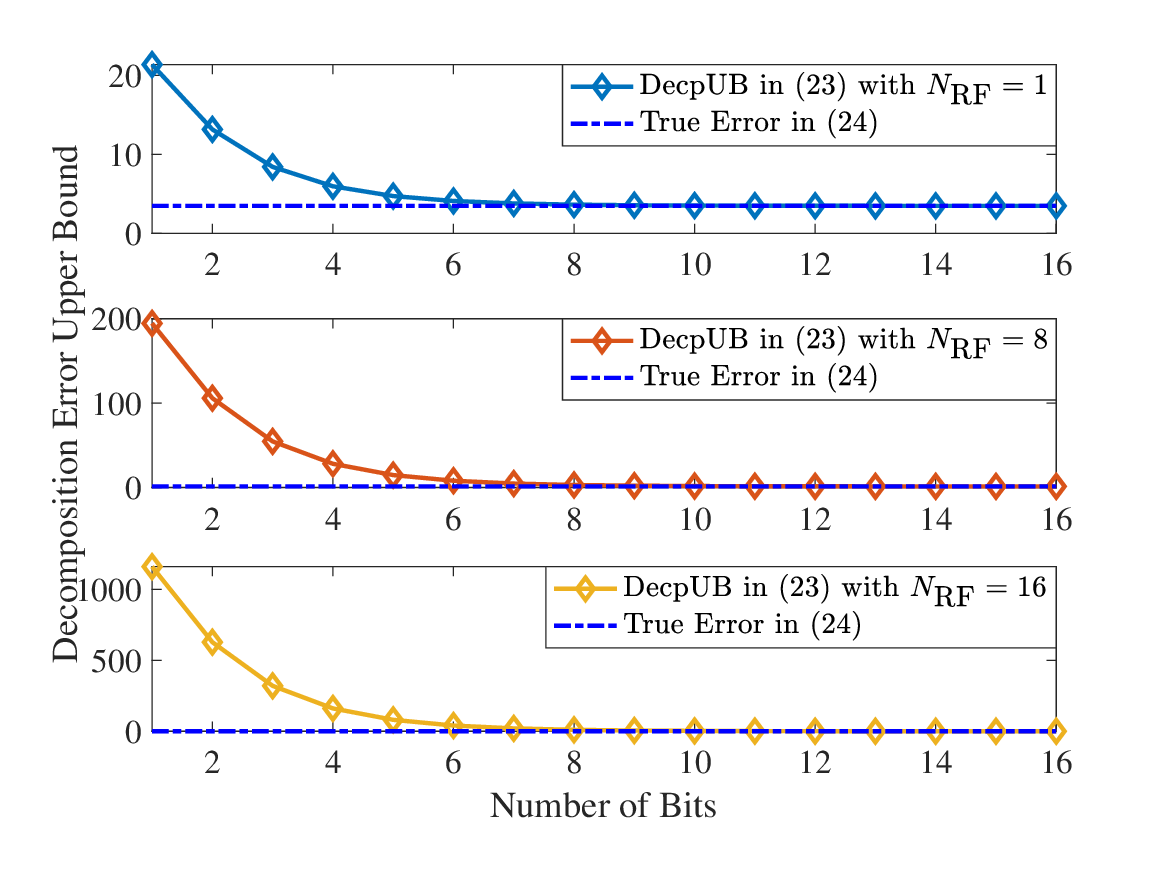}} 
    \vspace{-2mm}
	\caption{Decomposition error upper bound versus number of quantization bits of phase shifter.}
	\label{DecpUB}
\end{figure}

Note that $\sqrt{N_{\text{Tx}}N_{\text{RF}}} \|{\bf F}_{\text{RF}}^{\star}\|_{\text{F}} \|{\bf F}_{\text{BB}}\|_{\text{F}}$ in \eqref{QuantizationIn_f} is a constant w.r.t. the number of quantization bits of phase shifters. For notational simplicity, we define $C \triangleq \sqrt{N_{\text{Tx}}N_{\text{RF}}} \|{\bf F}_{\text{RF}}^{\star}\|_{\text{F}} \|{\bf F}_{\text{BB}}\|_{\text{F}}$, and rewrite the quantization error as
\begin{align*}
    \|{\bf F}_{\text{opt}} - {\bf{\widehat{F}}}_{\text{RF}}{\bf F}_{\text{BB}}\|_{\text{F}} - \|{\bf F}_{\text{opt}} - {\bf F}_{\text{RF}}^{\star}{\bf F}_{\text{BB}}\|_{\text{F}} \leq C \left|1 - e^{\jmath {\pi}/{2^{B}}} \right|.
\end{align*}
The values of $\left|1 - e^{\jmath {\pi}/{2^{B}}} \right|$ with different numbers of quantization bits are presented in Table~\ref{tab:QuantazationError}. It is seen from Table~\ref{tab:QuantazationError} that when $B \geq 5$, the quantization upper bound decreases by more than 10 times compared to $B = 1$, suggesting that $B=5$ can be sufficient to approach the performance of infinite-resolution phase shifters. In order to illustrate the impact of $B$ on the decomposition error upper bound (DecpUB), we define
\begin{align}
    {\text{DecpUB}} \triangleq \|{\bf F}_{\text{opt}} - {\bf F}_{\text{RF}}^{\star}{\bf F}_{\text{BB}}\|_{\text{F}} + C \left|1 - e^{\jmath {\pi}/{2^{B}}} \right|,
\end{align}
and then plot it w.r.t. the number of quantization bits in Fig.~\ref{DecpUB}, where the simulation parameters are $N_{\text{Tx}} = 16$, $M = 20$, and $P = 10$ dBm. The curve labeled ``True Error'' denotes the decomposition error by using infinite-resolution phase shifters, that is 
\begin{align}
    \text{True Error} \triangleq \|{\bf F}_{\text{opt}} - {\bf F}_{\text{RF}}^{\star}{\bf F}_{\text{BB}}\|_{\text{F}}.
\end{align} 
It can be observed from Fig.~\ref{DecpUB} that when $B \leq 3$ the DecpUB decrease sharply, and when $B \geq 5$ the slopes of the DecpUB are approximately equal to 0 and the DecpUB is very close to the true error.\footnote{The difference between two curves in Fig.~\ref{DecpUB} is $C \left|1 - e^{\jmath {\pi}/{2^{B}}} \right|$, which decrease rapidly w.r.t. the number of bits $B$. For example, when $B = 8$, the difference is proportional to $\left|1 - e^{\jmath {\pi}/{2^{8}}} \right| \approx 0.0123$ (note that $C$ is a limited number); when $B$ increases, this factor decreases.} This leads to the same conclusion as the one drawn from Table~\ref{tab:QuantazationError}, that is, $B = 5$ is sufficient to approach the performance of infinite-resolution phase shifters. This will be further verified through simulations in Section \ref{NumericalResults}.

\begin{table}[t]
\caption{Quantization error bound with different $B$}
\label{tab:QuantazationError}
\begin{tabular}{|c|c|c|c|c|c|c|}
    \hline
    \multirow{2}{*}{$B$} & \multirow{2}{*}{1} & \multirow{2}{*}{2} & \multirow{2}{*}{3} & \multirow{2}{*}{4} & \multirow{2}{*}{5} & \multirow{2}{*}{6} \\
    & & & & & & \\
    \hline
    \multirow{2}{*}{$\left|1 - e^{\jmath {\pi}/{2^{B}}} \right|$} & \multirow{2}{*}{1.4142} & \multirow{2}{*}{0.7654} & \multirow{2}{*}{0.3902} & \multirow{2}{*}{0.1960} & \multirow{2}{*}{0.0981} & \multirow{2}{*}{0.0491} \\
    & & & & & & \\
    \hline
\end{tabular}
\end{table}

\subsection{Convergence Analysis}
\label{ConvergenceAnalysis}
In this subsection, we analyse the convergence behaviors of the proposed ADMM algorithm (i.e., the inner iteration of Algorithm \ref{Proposed_Alg}) and AltOpt-LS-ADMM algorithm (i.e., the outer iteration of Algorithm \ref{Proposed_Alg}), which are stated in the following two theorems, respectively.
\begin{theorem}
\label{ADMM_convergence_theorem}
    The augmented Lagrangian function value sequence $\left\{\mathcal{L}\left( {\bf{\tilde F}}_{\text{\emph{RF}}}^{(k)}, {\bf F}_{\text{\emph{RF}}}^{(k)}, {\bf U}^{(k)} \right)  \Big| k = 0 , 1, 2, \cdots \right\}$ produced by the proposed ADMM algorithm converges if
    \begin{align}
    \label{rho_convergent_condition}
        \rho \geq \mathrm{max} \left\{ \sqrt{2}\|{\bf F}_{\text{\emph{BB}}}{\bf F}_{\text{\emph{BB}}}^{\textrm{\emph{H}}}\|_{\text{\emph{F}}} , \|{\bf F}_{\text{\emph{BB}}}\|_{\text{\emph{F}}}^{2} \right\}.
    \end{align}
    Furthermore, as $k \to \infty$, we have ${\bf F}_{\text{\emph{RF}}}^{(k+1)} = {\bf F}_{\text{\emph{RF}}}^{(k)}$, ${\bf{\tilde F}}_{\text{\emph{RF}}}^{(k+1)} = {\bf{\tilde F}}_{\text{\emph{RF}}}^{(k)}$, ${\bf U}^{(k+1)} = {\bf U}^{(k)}$, and ${\bf F}_{\text{\emph{RF}}}^{(k)} = {\bf{\tilde F}}_{\text{\emph{RF}}}^{(k)}$; and the point sequence $\left\{\left( {\bf{\tilde F}}_{\text{\emph{RF}}}^{(k)}, {\bf F}_{\text{\emph{RF}}}^{(k)}, {\bf U}^{(k)} \right)\right\}$ is a Cauchy sequence and it converges to a fixed point after a finite number of iterations.
\end{theorem}

\begin{proof}
    See Appendix \ref{ADMM_convergence_proof}.
\end{proof}

\begin{theorem}
\label{AltOptLSADMM_convergence_theorem}
    If \eqref{rho_convergent_condition} holds, the sequence $\left\{ \| {\bf F}_{\text{\emph{opt}}} - {\bf F}_{\text{\emph{RF}}}^{(i)}{\bf F}_{\text{\emph{BB}}}^{(i)} \|_{\text{\emph{F}}} \right\}$ generated by the proposed algorithm converges.
\end{theorem}

\begin{proof}
    See Appendix \ref{AltOptLSADMM_convergence_proof}.
\end{proof}

Theorem \ref{ADMM_convergence_theorem} asserts that as long as the augmented Lagrangian parameter $\rho$ is large enough (see \eqref{rho_convergent_condition}), the proposed ADMM algorithm is locally convergent. Additionally, Theorem \ref{AltOptLSADMM_convergence_theorem} establishes that the proposed AltOpt-LS-ADMM can generate a convergent sequence of cost function values, if \eqref{rho_convergent_condition} holds.

\section{Numerical Results}
\label{NumericalResults}

\subsection{Scenario, Performance Metric, and Benchmark}
In this section, we conduct simulations to verify the performance of the proposed AltOpt-LS-ADMM algorithm. For convenience, we consider a uniform linear array (ULA) at the BS, and its steering vector is given as
\begin{align*}
    {\bf a}(\theta) = \left[ 1, e^{- \jmath \frac{2\pi d}{\lambda}\sin{\theta} } , \cdots , e^{- \jmath \frac{2\pi d}{\lambda} (N_\text{Tx} - 1) \sin{\theta} } \right]^{\textrm{T}},
\end{align*}
with $d$ being the element spacing of the ULA and $\lambda$ denoting the transmit signal wavelength. Two simulation scenarios\footnote{\textbf{Scenario I} is to test whether the proposed Alt-LS-ADMM algorithm has satisfactory performance with any arbitrary precoder matrix, in terms of decomposition error.} are considered as follows:
\begin{itemize}
    \item \textbf{Scenario I:} We first test the performance with arbitrary ${\bf F}_{\text{opt}}$. We randomly generate a digital precoder ${\bf F}_{\text{opt}}$, and our performance metric is the decomposition error (DecpErr), defined as: ${\|{\bf F}_{\text{opt}} - {\bf F}_{\text{RF}} {\bf F}_{\text{BB}}\|_{\text{F}}} / {\|{\bf F}_{\text{opt}}\|_{\text{F}}}$. The simulation parameters are summarized in Table~\ref{tab:simulationparameters_I}.
    \item \textbf{Scenario II:} We then test the performance with the ${\bf F}_{\text{opt}}$ obtained from \textbf{Step 1} in Section \ref{ProposedTwoStepStrategy} for a single UE (Section \ref{Simulation_Scenario2}) and for multiple UEs (Section \ref{Simulation_Scenario2-multiple}). Our performance metric is the AEB in \eqref{AEB}. The simulation parameters are summarized in Table~\ref{tab:simulationparameters_II}.
\end{itemize}

\noindent We compare the proposed method with the following methods:
\begin{itemize}
    \item Alt-Babai \cite{Lyu2021}: alternating optimization + the Babai algorithm
    \item Alt-CDM \cite{Chen2017Aug}: alternating optimization + coordinate descent method
    \item Spa-OMP \cite{Ayach2014}: spatially sparse representation + orthogonal matching pursuit
    \item ManiOpt \cite{Yu2016}: manifold optimization (where we utilize the \texttt{Manopt} function \cite{Boumal2014} for implementation)
\end{itemize}
Note that ManiOpt in \cite{Yu2016} utilizes infinite-resolution (i.e., $B = \infty$) phase shifters, which is adopted as a benchmark in this work. Also note that the ManiOpt is initialized with random value, with the output of the Alt-Babai algorithm, or with the output of the proposed AltOpt-LS-ADMM method, labelled as ``ManiOpt (random init.)'', ``ManiOpt (Alt-Babai init.)'' and ``ManiOpt (proposed init.)'', respectively.

\begin{table}[t]
\caption{Simulation parameters in Scenario I}
\label{tab:simulationparameters_I}
\centering
\begin{tabular}{|c|c|}
    \hline
    \textbf{Parameter} & \textbf{Scenario I}  \\ 
    \hline
    $I_{\text{max}}$ & $10$ \\
    \hline
    $k_{\text{max}}$ & $50$  \\
    \hline
    BS Transmit Power, $P$  & $10$ dBm  \\
    \hline
    Pilot Length, $M$  & $20$  \\
    \hline
    $N_{\text{Tx}}$  & $16$  \\
    \hline
    \multirow{3}{*}{$N_{\text{RF}}$}  & $1, 2, \cdots, 16$ (Figs.~\ref{ErrovsRF} and \ref{ErrovsRF_compare}); \\  & $1,2,4,7,10,13,16$ (Fig.~\ref{ErrvsBits}); \\  &  $8$ (Fig.~\ref{ErrovsBits_compare})  \\
    \hline
    \multirow{3}{*}{$B$}  &  $1,2,\cdots,16$ (Figs.~\ref{ErrvsBits} and \ref{ErrovsBits_compare}); \\ & $1,2,3,4,5, \infty$ (Fig.~\ref{ErrovsRF}); \\  & 2 (Fig.~\ref{ErrovsRF_compare})  \\
    \hline
\end{tabular}
\end{table}

\begin{table}[t]
\vspace{2mm}
\caption{Simulation parameters in Scenario II}
\label{tab:simulationparameters_II}
\centering
\begin{tabular}{|c|c|}
    \hline
    \textbf{Parameter} & \textbf{Scenario II}  \\ 
    \hline
    $I_{\text{max}}$ & $10$ \\
    \hline
    $k_{\text{max}}$ & $50$  \\
    \hline
    BS Transmit Power, $P$  & $10$ dBm  \\
    \hline
    SNR   &  $10$ dB  \\
    \hline
    Pilot Length, $M$  & $20$  \\
    \hline
    \multirow{3}{*}{$N_{\text{Tx}}$}  & $8, 16, \cdots, 80$ (Figs.~\ref{AEBvsTx} and \ref{TimevsTxRF}); \\  &  $20$ (Figs.~\ref{AEBvsRF}, \ref{AEBvsBit}, \ref{AEBvsAoD}, and \ref{AEBvsRF_multipleUE});  \\
     & $4, 6, \cdots, 30$ (Fig.~\ref{AEBvsTx_multipleUE}) \\
    \hline
    \multirow{2}{*}{$N_{\text{RF}}$}  & $1, 2, \cdots, 20$ (Figs.~\ref{AEBvsRF}, \ref{TimevsTxRF}, and \ref{AEBvsRF_multipleUE}); \\  &  $4$ (Figs.~\ref{AEBvsTx}, \ref{AEBvsBit}, \ref{AEBvsAoD}, and \ref{AEBvsTx_multipleUE})  \\
    \hline
    \multirow{2}{*}{$B$}  & $\infty$ (Figs.~\ref{AEBvsTx}, \ref{AEBvsRF}, \ref{AEBvsAoD}, \ref{TimevsTxRF}, \ref{AEBvsTx_multipleUE}, and \ref{AEBvsRF_multipleUE});  \\  &  $1, 2, \cdots, 16$ (Fig.~\ref{AEBvsBit})  \\
    \hline
    \multirow{3}{*}{AoD}  & $-80^{\circ}, -75^{\circ}, \cdots, 80^{\circ}$ (Fig.~\ref{AEBvsAoD}); \\
     & $0^{\circ}$ and $60^{\circ}$ (Figs.~\ref{AEBvsTx_multipleUE} and \ref{AEBvsRF_multipleUE})  \\
    &  $0^{\circ}$ (Figs.~\ref{AEBvsTx}, \ref{AEBvsRF}, \ref{AEBvsBit}, and \ref{TimevsTxRF})  \\
    \hline
\end{tabular}
\end{table}

\subsection{Results and Discussion of Scenario I}

\subsubsection{DecpErr as a Function of $N_{\text{\emph{RF}}}$} 
\begin{figure}[t]
	\vspace*{-2mm}
	\centerline{\includegraphics[width=0.5\textwidth]{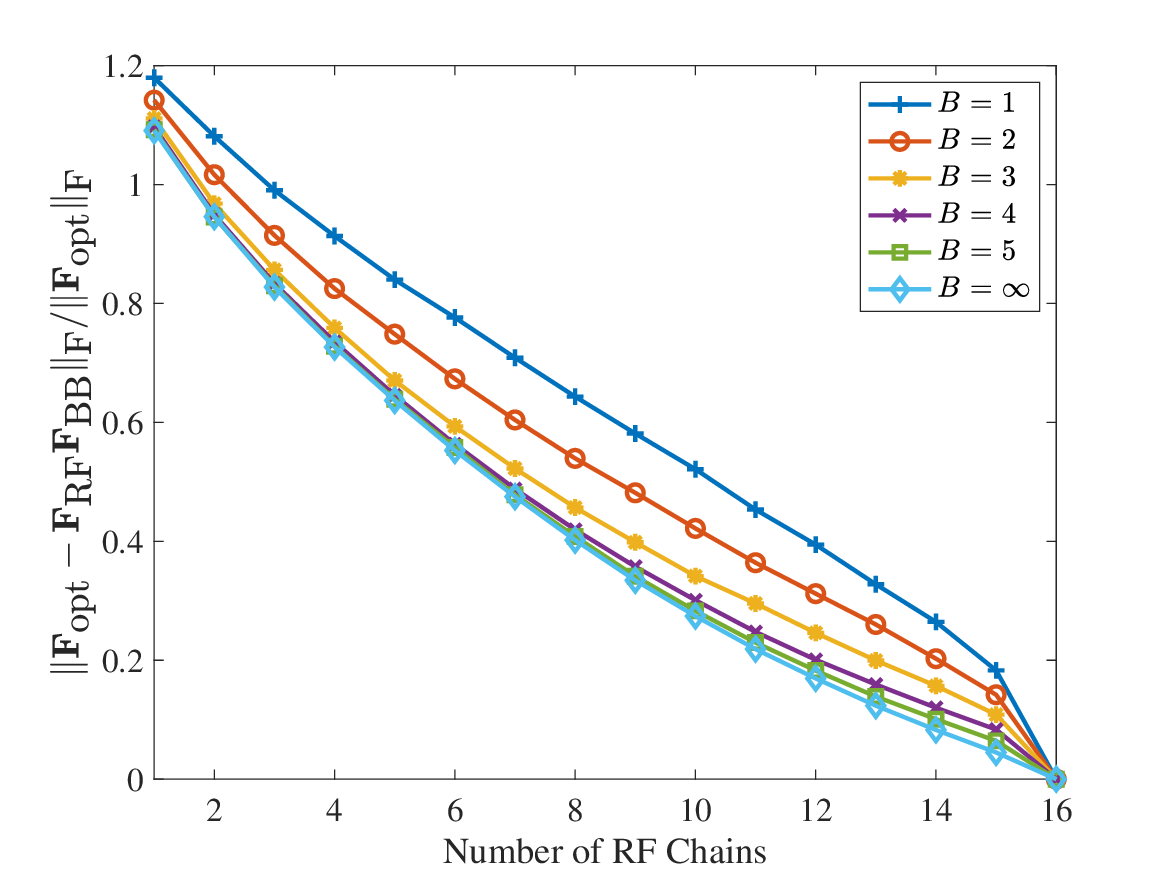}}  
	\caption{Decomposition error versus number of RF chains with different bits of the phase shifter, by the proposed method.}
	\label{ErrovsRF}
\end{figure}

We randomly\footnote{The entries of ${\bf F}_{\text{opt}}$ are first generated by drawing from the independent and identically distributed normal distribution, and then normalized to meet the power constraint $\|{\bf F}_{\text{opt}}\|_{\text{F}}^{2} = P$.} generate an ${\bf F}_{\text{opt}}$ and decompose it into ${\bf F}_{\text{RF}}$ and ${\bf F}_{\text{BB}}$ by using the proposed AltOpt-LS-ADMM algorithm. The DecpErrs are averaged over 500 Monte-Carlo trials, and the results w.r.t. the number of RF chains, $N_{\text{RF}}$, are plotted in Fig.~\ref{ErrovsRF}. It can be seen that: \textit{(i)} when $N_{\text{RF}}$ or $B$ increases, the DecpErr decreases; \textit{(ii)} when $B = 5$, its performance approaches the one with $B = \infty$ (i.e., the infinite resolution phase shifter); \textit{(iii)} when $N_{\text{RF}} = N_{\text{Tx}} = 16$, the DecpErrs are always 0. This is because when $N_{\text{RF}} = N_{\text{Tx}}$, ${\bf F}_{\text{RF}}$ is a square matrix and invertible, and thus there always exists a matrix ${\bf F}_{\text{BB}} = {\bf F}_{\text{RF}}^{-1}{\bf F}_{\text{opt}}$ such that ${\bf F}_{\text{opt}} = {\bf F}_{\text{RF}}{\bf F}_{\text{BB}}$.

\begin{figure}[t]
	\vspace*{-2mm}
	\centerline{\includegraphics[width=0.5\textwidth]{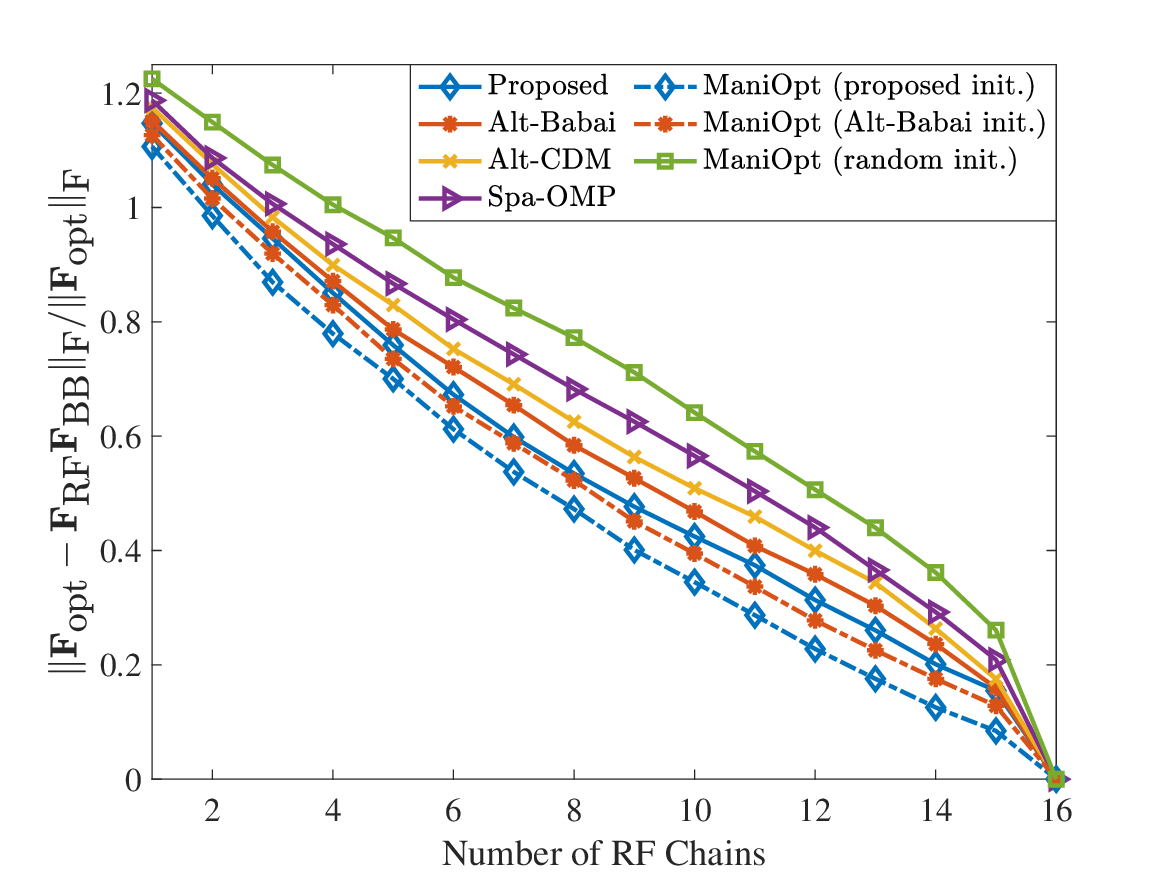}}  
	\caption{Decomposition error versus number of RF chains with $B = 2$ bits, among different algorithms.}
	\label{ErrovsRF_compare}
\end{figure}

Next, the DecpErrs of different methods are displayed in Fig.~\ref{ErrovsRF_compare}, with $B = 2$. We see that ManiOpt with random initialization has the worst performance and ManiOpt with the proposed method as initialization achieves the highest performance. Besides, the decomposition error of the proposed method is smaller than those of Alt-Babai, Alt-CDM, Spa-OMP, and ManiOpt with random initialization; and the decomposition error of ManiOpt (proposed init.) is smaller than that of ManiOpt (Alt-Babi init.). Note that ManiOpt with proposed initialization (or Alt-Babi initialization) attains the best (or the second best) decomposition performance at a cost of higher computational complexity, which will be verified in Fig.~\ref{TimevsTxRF}.

\subsubsection{DecpErr as a Function of $B$} 
\begin{figure}[t]
	\vspace*{-2mm}
	\centerline{\includegraphics[width=0.5\textwidth]{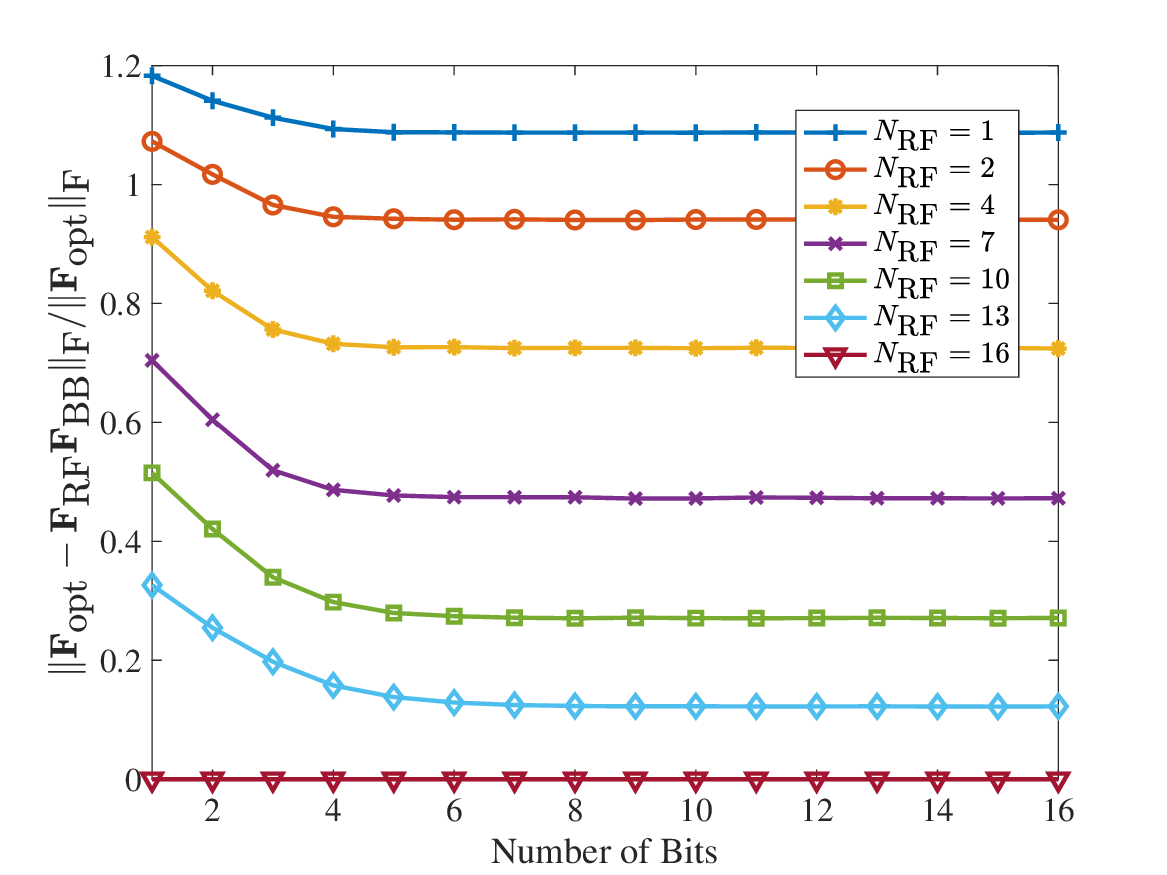}}  
	\caption{Decomposition error versus number of bits of the phase shifter with different numbers of RF chains, by the proposed method.}
	\label{ErrvsBits}
\end{figure}

The DecpErr of the proposed algorithm w.r.t. the number of quantization bits, $B$, is shown in Fig.~\ref{ErrvsBits}. We observe that when $N_{\text{RF}}$ increases, the decomposition error decreases, as expected. For $N_{\text{RF}} < 16$, when $B$ increases from 1 until 5, the decomposition error decreases; when $B \geq 5$, the decomposition error remains nearly unchanged. Therefore, taking into account the outcomes presented in Fig.~\ref{ErrovsRF}, we can infer that $B = 5$ bits prove to be sufficient in achieving near-optimal hybrid precoding performance (i.e., reaching a performance level very close to that obtained by digital precoding). Besides, it is seen from Fig.~\ref{ErrvsBits} that, the decomposition error is 0 when $N_{\text{RF}} = N_{\text{Tx}} = 16$, which has been explained in the first example.

\begin{figure}[t]
	\vspace*{-2mm}
	\centerline{\includegraphics[width=0.5\textwidth]{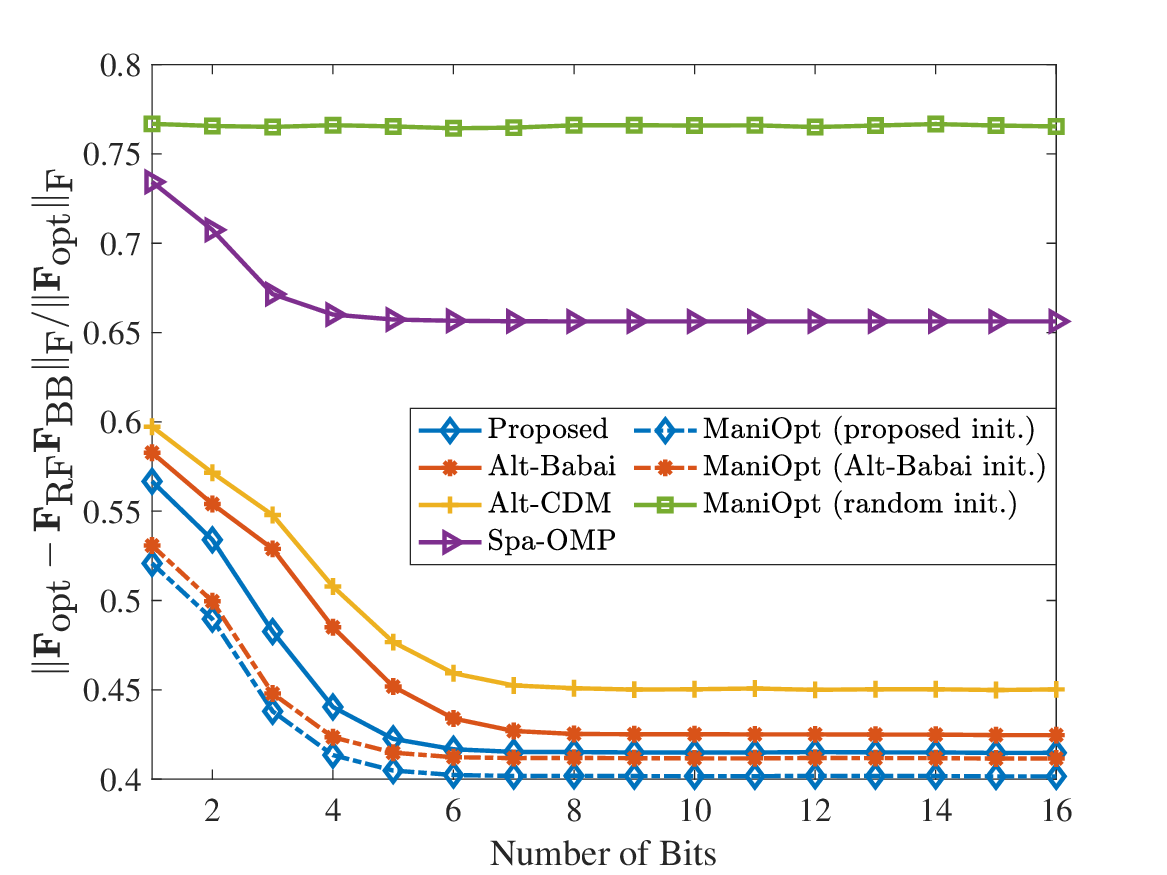}}  
	\caption{Decomposition error versus number of bits of the phase shifter with $N_{\text{RF}} = 8$ RF chains, among different algorithms.}
	\label{ErrovsBits_compare}
\end{figure}

Next, the DecpErrs of different algorithms are depicted in Fig.~\ref{ErrovsBits_compare}, with $N_{\text{RF}} = 8$, which verifies the better performance of the proposed method against Alt-Babai, Alt-CDM, Spa-OMP, and ManiOpt with random initialization. Note that the ManiOpt with random initialization has a horizontal line because it uses $B = \infty$ quantization bits; while ManiOpt (proposed init.) and ManiOpt (Alt-Babai) init. do not has a horizontal line because the manifold optimization method is sensitive to its initialization. ManiOpt (proposed init.) has lower decomposition error than ManiOpt (Alt-Babai init.), which indicates that the proposed algorithm produces a better output as an initialization for the ManiOpt.

\subsection{Results and Discussion of Scenario II -- Single UE}
\label{Simulation_Scenario2}

We first consider one UE at $\theta = 0^{\circ}$.

\subsubsection{AEB as a Function of $N_{\text{\emph{Tx}}}$} 
\begin{figure}[t]
	\vspace*{-5mm}
	\centerline{\includegraphics[width=0.5\textwidth]{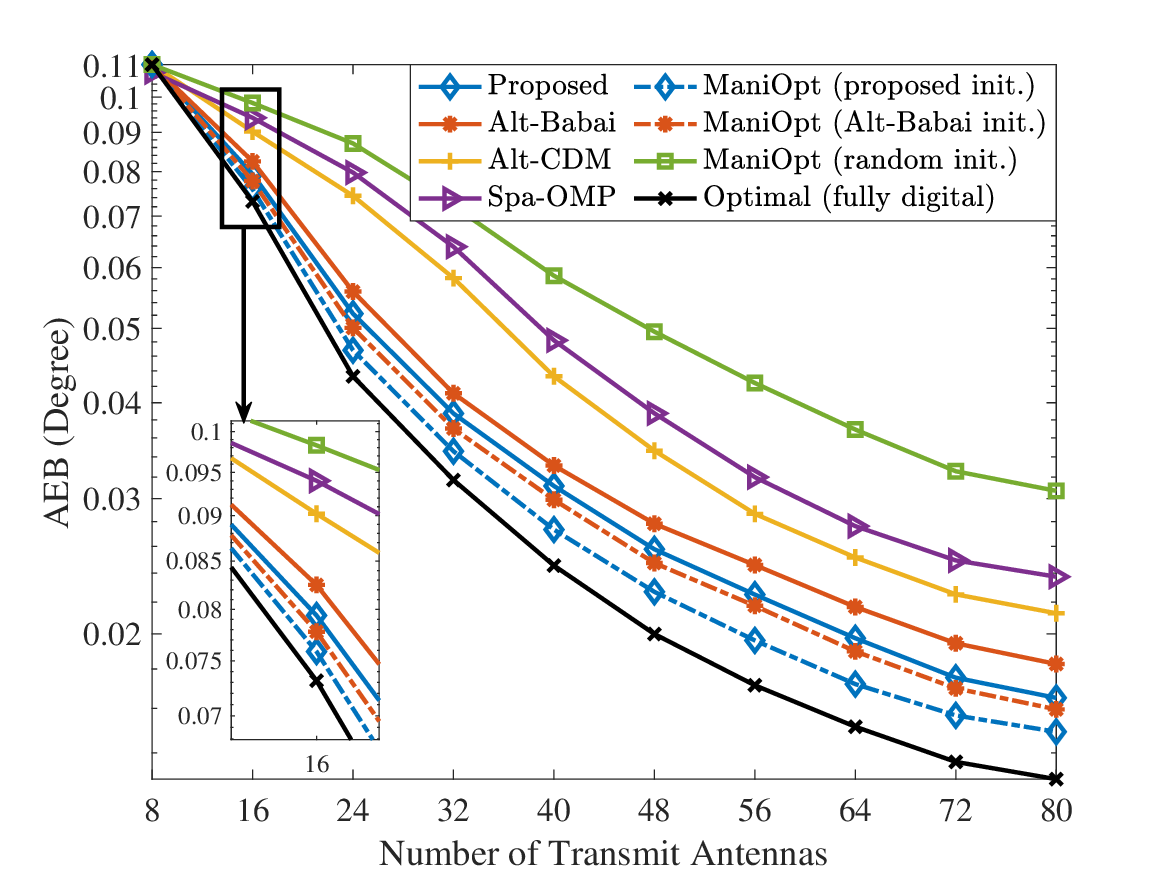}}   
	\caption{AEB versus number of transmit antennas with $N_{\text{RF}} \!=\! 8$, $B = \infty$, and SNR = 10 dB.}
	\label{AEBvsTx}
\end{figure}

We now evaluate the AEB performance of the proposed algorithm and the benchmark methods. The approximately optimal digital precoder ${\bf F}_{\text{opt}}$ can be obtained by the method as introduced in Section \ref{ProposedTwoStepStrategy}. Then we decompose this ${\bf F}_{\text{opt}}$ into ${\bf F}_{\text{RF}}$ and ${\bf F}_{\text{BB}}$ by using different algorithms. The AEBs w.r.t. the number of Tx antennas, $N_{\text{Tx}}$, achieved by the different methods are shown in Fig.~\ref{AEBvsTx}, where the curve labelled as ``Optimal (fully digital)'' is the result by using ${\bf F}_{\text{opt}}$ directly without decomposition. We can see that the proposed algorithm attains lower AEB than those of Alt-Babai, Alt-CDM, Spa-OMP, and ManiOpt with random initialization. The ManiOpt (proposed init.) outperforms the ManiOpt (Alt-Babai init.) and is much closer to the ``Optimal (fully digital)'' curve.

\subsubsection{AEB as a Function of $N_{\text{\emph{RF}}}$} 
\begin{figure}[t]
	\vspace*{-3mm}
	\centerline{\includegraphics[width=0.5\textwidth]{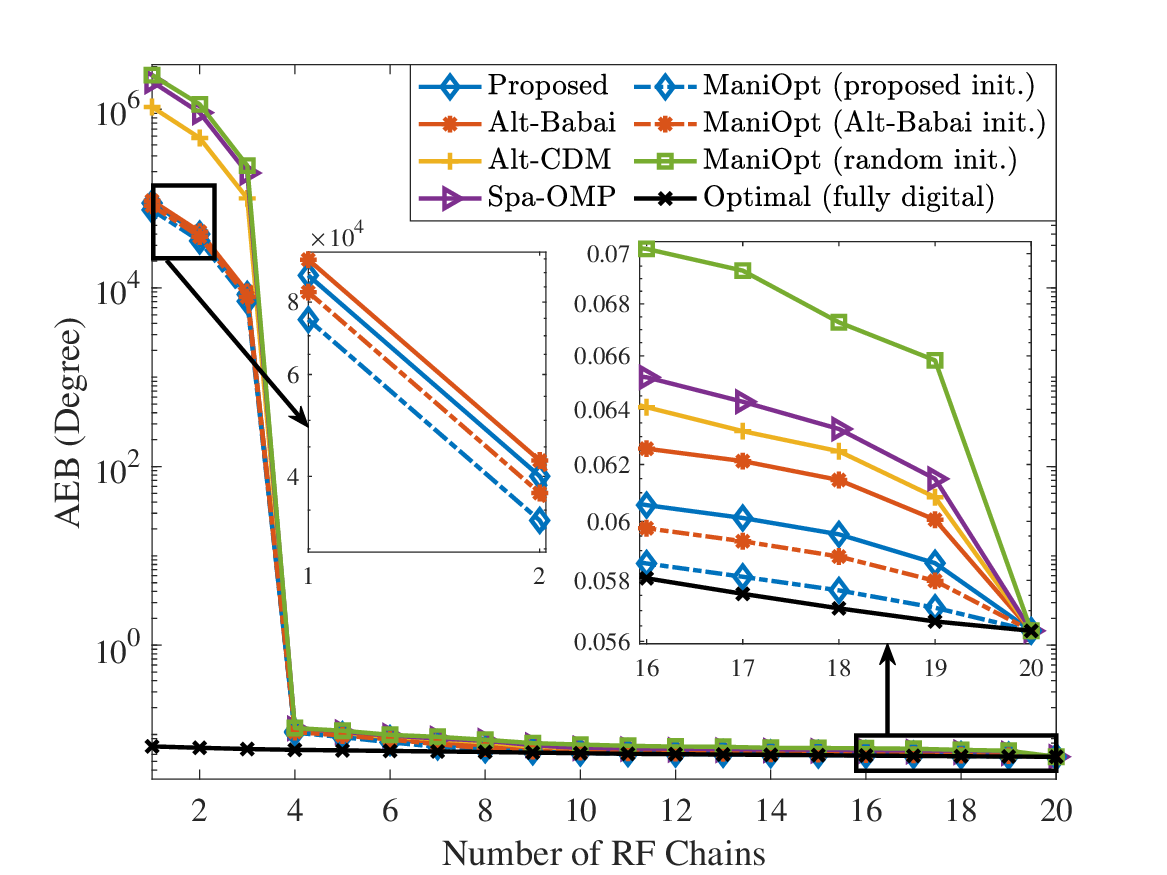}}   
	\caption{AEB versus number of RF chains with $N_{\text{Tx}} = 20$, $B = \infty$, and SNR = 10 dB.}
	\label{AEBvsRF}
\end{figure}

The results of AEB w.r.t. the number of RF chains, $N_{\text{RF}}$, are presented in Fig.~\ref{AEBvsRF}. It can be observed from Fig.~\ref{AEBvsRF} that the AEB of the proposed algorithm is smaller than those of Alt-Babai, Alt-CDM, Spa-OMP, and ManiOpt (random init.); while ManiOpt (proposed init.) outperforms ManiOpt (Alt-Babai init.) and has the best performance.

\subsubsection{AEB as a Function of $B$} 
\begin{figure}[t]
	\vspace*{-5mm}
	\centerline{\includegraphics[width=0.5\textwidth]{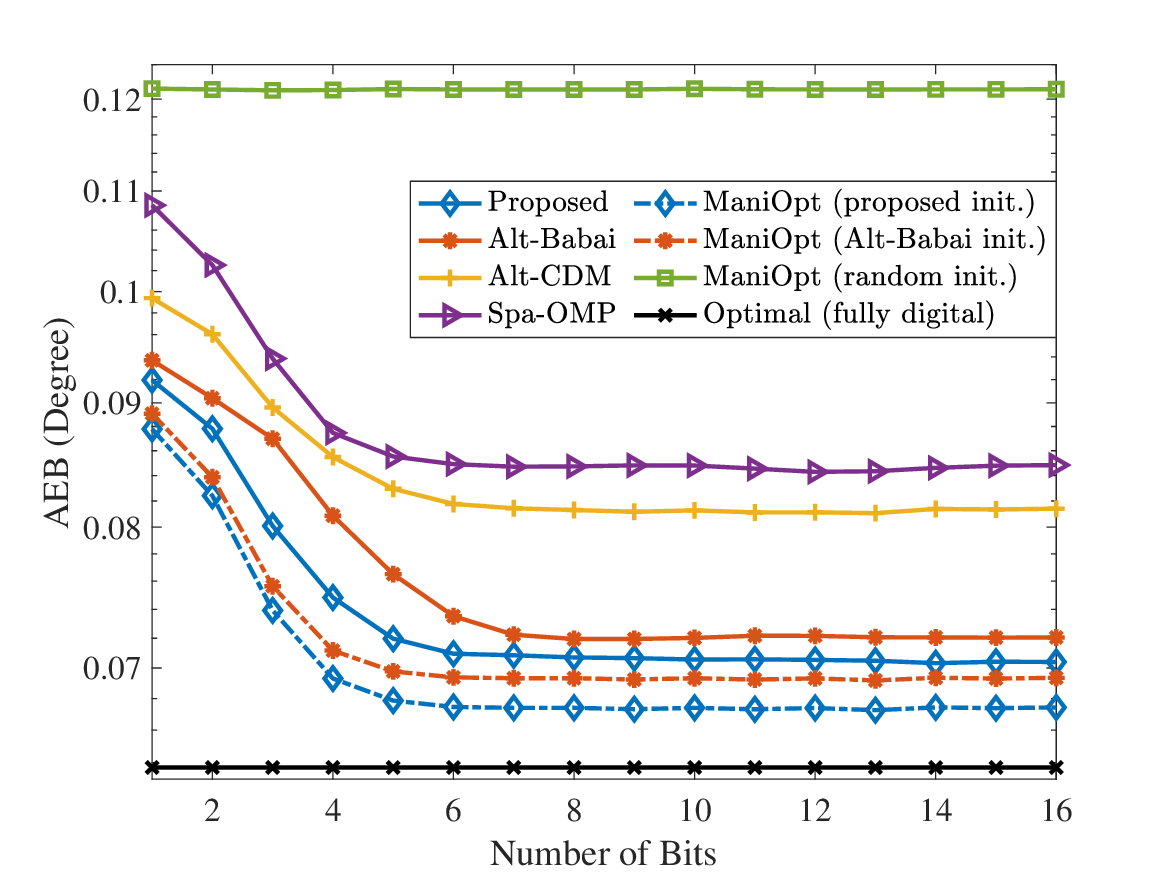}}   
	\caption{AEB versus number of bits of the phase shifters with $N_{\text{Tx}} = 20$, $N_{\text{RF}} = 8$, and SNR = 10 dB.}
	\label{AEBvsBit}
\end{figure}

We compare AEB versus different quantization bits of phase shifters, and the results are displayed in Fig.~\ref{AEBvsBit}. We have similar findings as in Fig.~\ref{ErrovsBits_compare}.

\subsubsection{AEB as a Function of AoD} 
\begin{figure}[t]
	\vspace*{-3mm}
	\centerline{\includegraphics[width=0.5\textwidth]{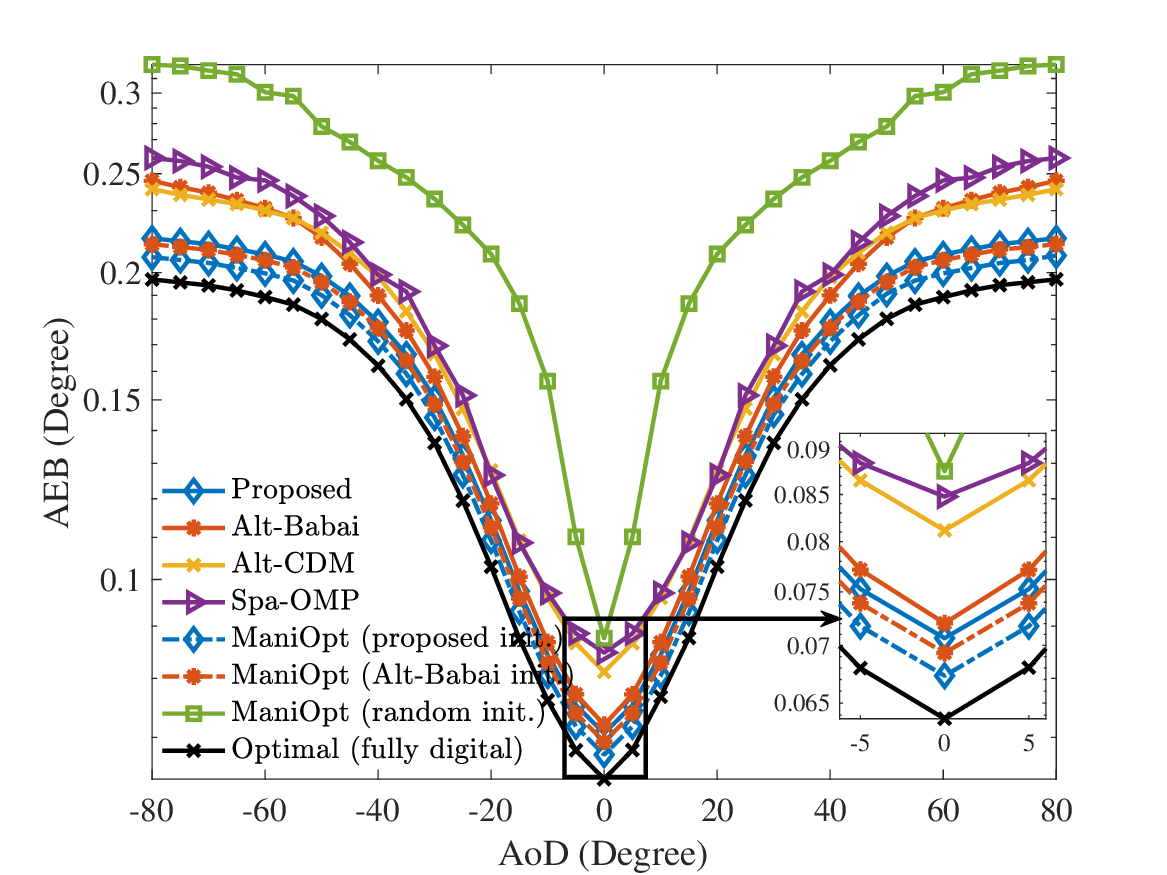}}   
	\caption{AEB versus AoD with $N_{\text{Tx}} = 20$, $N_{\text{RF}} = 8$, $B = \infty$, and SNR = 10 dB.}
	\label{AEBvsAoD}
\end{figure}

The predefined codebook for ${\bf F}_{\text{opt}}$ is set around\footnote{The predefined angles are chosen uniformly from $[-5^{\circ} , 5^{\circ})$ with step size $10^{\circ}/G$.} $0^{\circ}$, while the true AoD changes from $-80^{\circ}$ to $80^{\circ}$. The results of AEB versus AoD are plotted in Fig.~\ref{AEBvsAoD}. We see that when AoD is $0^{\circ}$ (matching with our predefined codebook), all the curves reach their lowest AEB. Besides, the ManiOpt (proposed init.) outperforms others, followed by ManiOpt (Alt-Babai init.) and then the proposed AltOpt-LS-ADMM algorithm. In addition, the proposed algorithm for hybrid precoder design, both when used independently and as an initialization for ManiOpt, exhibits AoD estimation performance very close to that achieved via a fully digital array, which underscores its effectiveness over a broad range of AoD values.  

\subsubsection{CPU Runtime as a Function of $N_{\text{\emph{Tx}}}$ or $N_{\text{\emph{RF}}}$} 
\begin{figure}[t]
	\vspace*{-2mm}
\centerline{\includegraphics[width=0.5\textwidth]{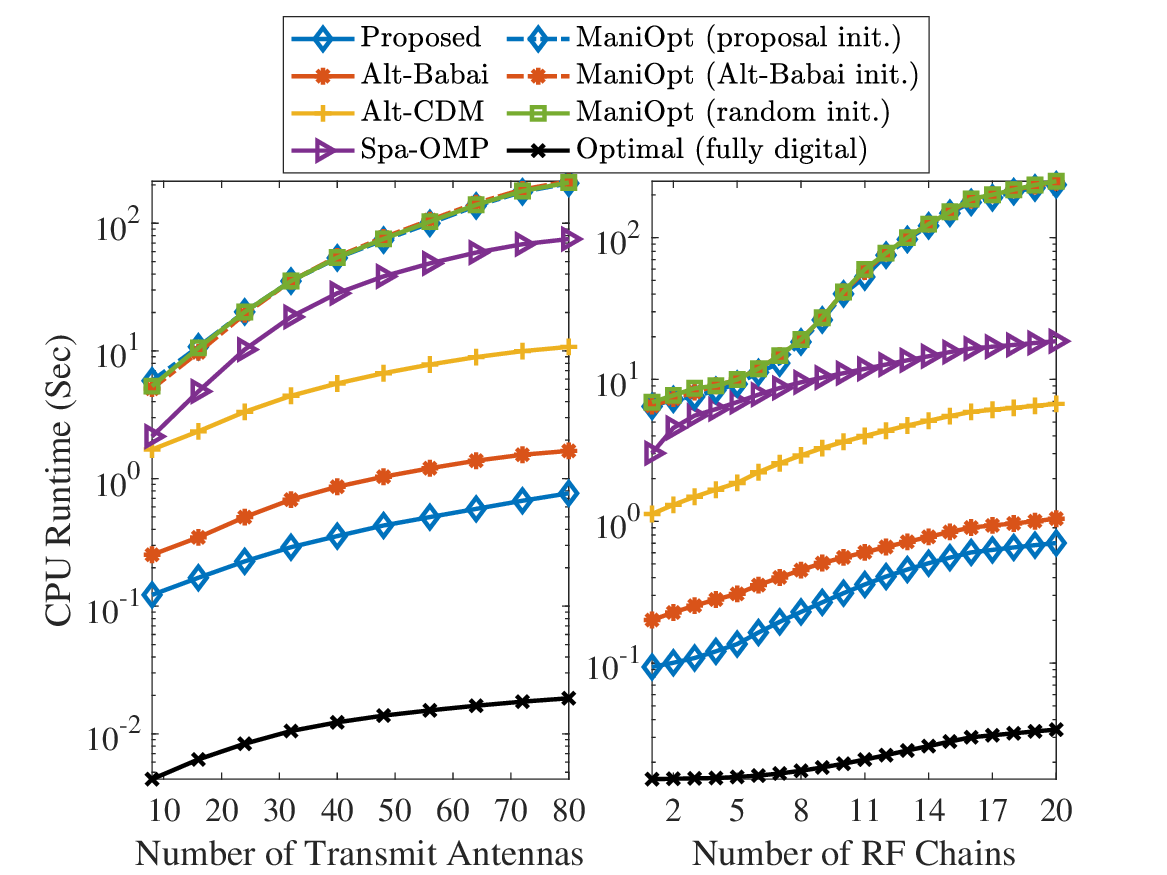}}   
	\caption{CPU runtime versus number of transmit antennas (left) and number of RF chains (right).}
	\label{TimevsTxRF}
\end{figure}

We compare the computational cost of different algorithms. The central processing unit (CPU) runtime versus number of transmit antennas is drawn in Fig.~\ref{TimevsTxRF} (left). Note that the CPU runtime for each method contains \textit{(i)} decomposing ${\bf F}_{\text{opt}}$ into ${\bf F}_{\text{RF}}$ and ${\bf F}_{\text{BB}}$ and \textit{(ii)} calculating the AEB via \eqref{AEB}. The ``Optimal (full digital)'' does not need \textit{(i)}. We show the CPU runtime of ``Optimal (full digital)'' as a benchmark. On the other hand, the CPU runtime versus number of RF chains is drawn in Fig.~\ref{TimevsTxRF} (right). From Fig.~\ref{TimevsTxRF}, it can be seen that ManiOpt with three different initializations have almost the same CPU runtime. The Optimal method has the least CPU runtime since it does not need to perform the decomposition operation on ${\bf F}_{\text{opt}}$. Besides, the proposed algorithm consumes less CPU runtime than Alt-Babai, Alt-CDM, Spa-OMP, and ManiOpt.

\subsection{Results and Discussion of Scenario II -- Multiple UEs}
\label{Simulation_Scenario2-multiple}

We now consider two UEs at $\theta_{1} = 0^{\circ}$ and $\theta_{2} = 60^{\circ}$, for the following three cases: \textit{(i)} The optimal fully digital precoder ${\bf F}_{\text{opt}}$ is designed to illuminate only the first UE. \textit{(ii)} The optimal fully digital precoder ${\bf F}_{\text{opt}}$ is designed to illuminate only the second UE. \textit{(iii)} The optimal fully digital precoder ${\bf F}_{\text{opt}}$ is designed to illuminate both UEs, as described in Remark \ref{remark_multipleUE}. In all cases, $M=20$ pilots are used.

\subsubsection{AEB as a Function of $N_{\text{\emph{Tx}}}$ with 2 UEs} 
\begin{figure}[t]
	\vspace*{-2mm}
	\centerline{\includegraphics[width=0.5\textwidth]{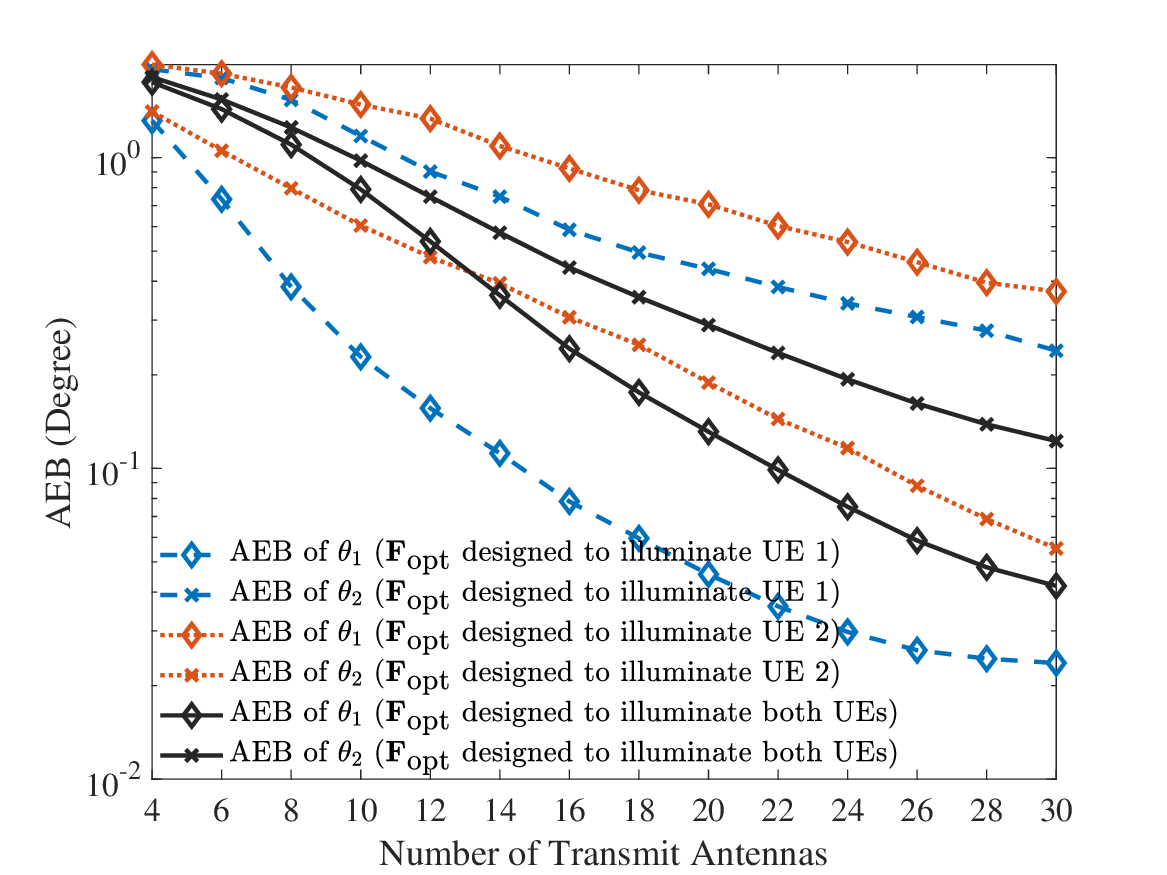}}
	\caption{AEB versus number of transmit antennas with $L = 2$ UEs at $\theta_{1} = 0^{\circ}$ and $\theta_{2} = 60^{\circ}$.}
	\label{AEBvsTx_multipleUE}
\end{figure}

The AEBs of $\theta_{1}$ and $\theta_{2}$ for each case w.r.t. number of transmit antennas are drawn in Fig.~\ref{AEBvsTx_multipleUE}. We see that if ${\bf F}_{\text{opt}}$ is designed to illuminate only UE 1 (resp. UE 2), the AEB of $\theta_{1}$ (resp. $\theta_{2}$) is smaller than those of cases with ${\bf F}_{\text{opt}}$ designed to illuminate only UE 2 (resp. UE 1) or both. Besides, the AEB of $\theta_{1}$ with ${\bf F}_{\text{opt}}$ designed to illuminate only UE 1 is smaller than the AEB of $\theta_{2}$ with ${\bf F}_{\text{opt}}$ designed to illuminate only UE 2. This is consistent with the result in Fig.~\ref{AEBvsAoD}. Further, ${\bf F}_{\text{opt}}$ designed to illuminate both UEs provides a good middle ground between the other two cases.

\subsubsection{AEB as a Function of $N_{\text{\emph{RF}}}$ with 2 UEs} 
\begin{figure}[t]
	\vspace*{-2mm}
	\centerline{\includegraphics[width=0.5\textwidth]{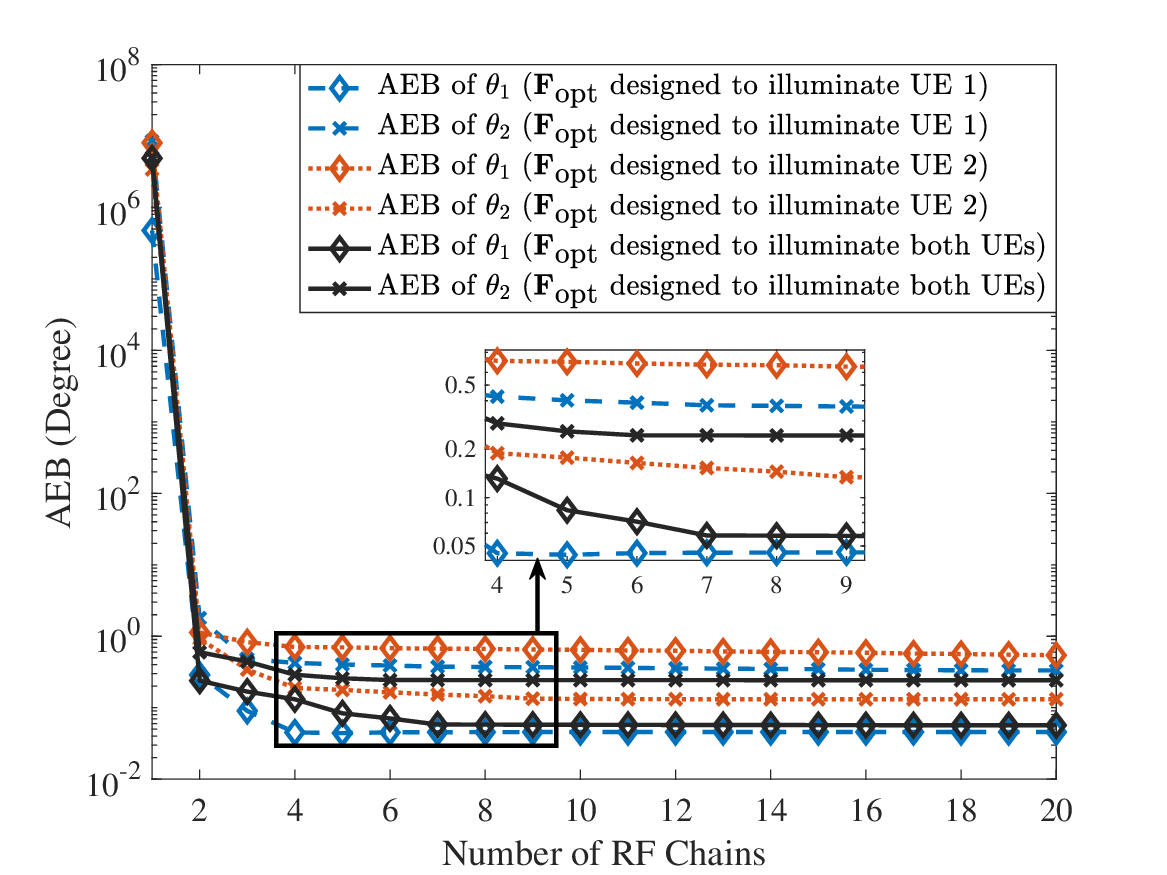}}
	\caption{AEB versus number of RF chains with $L = 2$ UEs at $\theta_{1} = 0^{\circ}$ and $\theta_{2} = 60^{\circ}$.}
	\label{AEBvsRF_multipleUE}
\end{figure}

The AEBs of $\theta_{1}$ and $\theta_{2}$ for each case w.r.t. number of RF chains are drawn in Fig.~\ref{AEBvsRF_multipleUE}. We observe similar results as the previous example.

\section{Conclusion}
\label{Conclusion}

In this paper, we have investigated the hybrid precoder design problem for angle-of-departure (AoD) estimation, where we took into account practical limitation on the finite resolution of phase shifters. Our aim was to devise a radio-frequency (RF) precoder and a base-band (BB) precoder that could simultaneously adhere to the practical constraint and achieve highly precise AoD estimation. To accomplish this goal, we developed a two-step approach. Firstly, we derived a fully digital precoder that minimizes the angle error bound by using a predefined codebook. Then, we decomposed this digital precoder into an RF precoder and a BB precoder, employing the alternating optimization framework and alternating direction method of multipliers. We also analysed the quantization error bound, and provided convergence analyses of the proposed algorithm. Numerical results demonstrated the exceptional performance of the proposed method with low complexity, leading to the following key conclusions: 
\begin{itemize}
\item \textit{Number of Bits Sufficient for AoD Estimation:} 5 bits are sufficient to achieve almost the same decomposition and AoD estimation performance as the case with infinite-resolution phase shifters. 
\item \textit{Number of RF Chains Sufficient for AoD Estimation:} For a 20-element transmit array, 4 RF chains are sufficient to attain the same AoD estimation performance as the fully-digital architecture.
\item \textit{High-Quality Initialization:} The proposed algorithm can provide high-quality initialization that improves the performance of manifold optimization compared to both random initialization and the initialization that uses the output of the best state of the art. 
\item \textit{Covering Wide Range of AoDs:} The proposed algorithm attains near-optimal (in the sense of achieving the fully-digital performance) AoD performance over a broad range of AoD values ranging from -80 to 80 degrees.
\end{itemize}

\appendices
\section{Calculation of the FIM}
\label{appendix_A}
The derivatives of ${\bf{\tilde{y}}} = \beta {\bf S}({\bf F}_{\text{RF}}{\bf F}_{\text{BB}})^{\textrm{T}}{\bf a}(\theta)$ w.r.t. $[{\bf x}]_{i}$, $i = 1, 2, 3$, are given as follows
\begingroup
\allowdisplaybreaks
\begin{subequations}
\begin{align*}
    \frac{\partial {\bf{\tilde{y}}}}{\partial [{\bf x}]_{1}} & = \frac{\partial {\bf{\tilde{y}}}}{\partial \sin{\theta}} = - \jmath \frac{2 \beta \pi d }{\lambda} {\bf S}({\bf F}_{\text{RF}}{\bf F}_{\text{BB}})^{\textrm{T}} {\bf D} {\bf a}(\theta),  \\
    \frac{\partial {\bf{\tilde{y}}}}{\partial [{\bf x}]_{2}} & = \frac{\partial {\bf{\tilde{y}}}}{\partial \beta_{\text{R}}} = {\bf S}({\bf F}_{\text{RF}}{\bf F}_{\text{BB}})^{\textrm{T}} {\bf a}(\theta),  \\
    \frac{\partial {\bf{\tilde{y}}}}{\partial [{\bf x}]_{3}} & = \frac{\partial {\bf{\tilde{y}}}}{\partial \beta_{\text{I}}} = \jmath {\bf S}({\bf F}_{\text{RF}}{\bf F}_{\text{BB}})^{\textrm{T}} {\bf a}(\theta),
\end{align*}
\end{subequations}
\endgroup
where ${\bf D} \triangleq \text{diag}\{0, 1, \cdots, N_{\text{Tx}-1}\}$. According to \eqref{FIM}, we have
\begingroup
\allowdisplaybreaks
\begin{subequations}
\begin{align*}
    [{\bf J}]_{11} & = \frac{8 |\beta|^{2}\pi^{2}d^{2} }{\sigma_{\text{n}}^{2} \lambda^{2}} {\bf a}^{\textrm{H}}(\theta){\bf D}{\bf F}^{*}{\bf S}^{\textrm{H}}{\bf S}{\bf F}^{\textrm{T}}{\bf D}{\bf a}(\theta), \\
    [{\bf J}]_{12} & = [{\bf J}]_{21} = \frac{4 \pi d}{ \sigma_{\text{n}}^{2} \lambda} \times \Re \left\{ \jmath \beta^{*} {\bf a}^{\textrm{H}}(\theta){\bf D}{\bf F}^{*}{\bf S}^{\textrm{H}}{\bf S}{\bf F}^{\textrm{T}}{\bf a}(\theta) \right\}, \\
    [{\bf J}]_{13} & = [{\bf J}]_{31} = - \frac{4 \pi d}{ \sigma_{\text{n}}^{2} \lambda} \times \Re \left\{ \beta^{*} {\bf a}^{\textrm{H}}(\theta){\bf D}{\bf F}^{*}{\bf S}^{\textrm{H}}{\bf S}{\bf F}^{\textrm{T}}{\bf a}(\theta) \right\}, \\
    [{\bf J}]_{22} & = [{\bf J}]_{33} = \frac{2}{\sigma_{\text{n}}^{2}} {\bf a}^{\textrm{H}}(\theta) {\bf F}^{*} {\bf S}^{\textrm{H}}{\bf S}{\bf F}^{\textrm{T}}{\bf a}(\theta), \\
    [{\bf J}]_{23} & = [{\bf J}]_{32} = 0.
\end{align*}
\end{subequations}
\endgroup

\section{Calculation of the AEB}
\label{appendix_B}
Based on the block matrix inversion lemma \cite{Anton1994}, we have
\begingroup
\allowdisplaybreaks
\begin{subequations}
\begin{align*}
    & \text{AEB}({\bf F}_{\text{RF}}, {\bf F}_{\text{BB}}; {\bf x}) \\
    = & ~\! \sqrt{[{\bf C}]_{11}} \\
    = & ~\! \sqrt{[{\bf J}^{-1}]_{11}}  \\
    = & ~\! \sqrt{\left[ \left[ \begin{array}{ccc}
      \left[{\bf J}\right]_{11} & \left[{\bf J}\right]_{12} & \left[{\bf J}\right]_{13}  \\
      \left[{\bf J}\right]_{21} & \left[{\bf J}\right]_{22} & \left[{\bf J}\right]_{23}  \\
      \left[{\bf J}\right]_{31} & \left[{\bf J}\right]_{32} & \left[{\bf J}\right]_{33}
    \end{array} \right]^{-1} \right]_{11} }   \\
    = & ~\! \sqrt{ \left( \! [{\bf J}]_{11} - \bigg[ [{\bf J}]_{12} , [{\bf J}]_{13} \bigg] \!\! \left[ \!\!\! \begin{array}{cc}
    [{\bf J}]_{22} & [{\bf J}]_{23} \\
    \left[ {\bf J} \right]_{32} & [{\bf J}]_{33}
    \end{array}
    \!\!\! \right]^{-1}
    \! \left[ \!\! \begin{array}{c}
    \left[ {\bf J} \right]_{21}  \\
    \left[ {\bf J} \right]_{31}
    \end{array}
    \!\! \right]
    \! \right)^{-1}}   \\
    = & ~\! \left( \! [{\bf J}]_{11} - \bigg[ [{\bf J}]_{12} , [{\bf J}]_{13} \bigg] \!\! \left[ \!\!\! \begin{array}{cc}
    [{\bf J}]_{22} & [{\bf J}]_{23} \\
    \left[ {\bf J} \right]_{32} & [{\bf J}]_{33}
    \end{array}
    \!\!\! \right]^{-1}
    \! \left[ \!\! \begin{array}{c}
    \left[ {\bf J} \right]_{21}  \\
    \left[ {\bf J} \right]_{31}
    \end{array}
    \!\! \right]
    \! \right)^{-\frac{1}{2}}  \\
    = & ~\! \left( \! [{\bf J}]_{11} - \bigg[ [{\bf J}]_{12} , [{\bf J}]_{13} \bigg] \!\! \left[ \!\!\! \begin{array}{cc}
    [{\bf J}]_{22} & 0 \\
    0 & [{\bf J}]_{22}
    \end{array}
    \!\!\! \right]^{-1}
    \! \left[ \!\! \begin{array}{c}
    \left[ {\bf J} \right]_{12}  \\
    \left[ {\bf J} \right]_{13}
    \end{array}
    \!\! \right]
    \! \right)^{-\frac{1}{2}}   \\
    = & ~\! \left( [{\bf J}]_{11} - \frac{ (\left[{\bf J}\right]_{12})^{2} + ([{\bf J}]_{13})^{2} }{[{\bf J}]_{22}} \right)^{-\frac{1}{2}}.
\end{align*} 
\end{subequations}
\endgroup
Substituting the results in Appendix \ref{appendix_A} and ${\bf S}^{\textrm{H}}{\bf S} = \sigma_{\text{s}}^{2} {\bf I}_{M}$ into the above equation yields (\ref{AEB}).

\section{Proof of Theorem \ref{ADMM_convergence_theorem}}
\label{ADMM_convergence_proof}
To show the convergence of $\left\{ \mathcal{L} \! \left({\bf{\tilde F}}_{\text{RF}}^{(k)}, {\bf F}_{\text{RF}}^{(k)}, {\bf U}^{(k)}\right) \right\}$, we first provide the following two lemmas:
\begin{lemma}
\label{monotonicity_lemma}
    The proposed AltOpt-LS-ADMM algorithm, i.e., Algorithm \ref{Proposed_Alg}, produces a monotonically decreasing sequence $\{\mathcal{L}^{(k)}  | k = 0 , 1, 2, \cdots \}$, where $\mathcal{L}^{(k)} \triangleq \mathcal{L}\left({\bf{\tilde F}}_{\text{\emph{RF}}}^{(k)}, {\bf F}_{\text{\emph{RF}}}^{(k)}, {\bf U}^{(k)}\right)$, provided that the augmented Lagrangian parameter $\rho$ satisfies
    \begin{align}
        \rho \geq \sqrt{2}\| {\bf F}_{\text{\emph{BB}}}{\bf F}_{\text{\emph{BB}}}^{\textrm{\emph{H}}} \|_{\text{\emph{F}}}.
    \end{align}
\end{lemma}
\begin{lemma}
\label{boundedness_lemma}
    The function $\mathcal{L}\left({\bf{\tilde F}}_{\text{\emph{RF}}}, {\bf F}_{\text{\emph{RF}}}, {\bf U}\right)$ defined in \eqref{AugLagFun} is bounded from below by 0 during the iteration process \eqref{Update_F_U}, provided that the augmented Lagrangian parameter $\rho$ satisfies
    \begin{align}
        \rho \geq \|{\bf F}_{\text{\emph{BB}}}\|_{\text{\emph{F}}}^{2}.
    \end{align}
\end{lemma}
\noindent The proofs of Lemmas~\ref{monotonicity_lemma} and \ref{boundedness_lemma} are relegated to Appendix~\ref{monotonicity_proof} and Appendix~\ref{boundedness_proof}, respectively. These two lemmas straightforwardly implies that the sequence $\left\{ \mathcal{L} \! \left({\bf{\tilde F}}_{\text{RF}}^{(k)}, {\bf F}_{\text{RF}}^{(k)}, {\bf U}^{(k)}\right) \right\}$ is convergent. Therefore, when the augmented Lagrangian parameter $\rho$ satisfies \eqref{rho_convergent_condition}, we have
\begin{align}
\label{L_diff_0}
    \mathcal{L}^{(k+1)} - \mathcal{L}^{(k)} = 0
\end{align}
as $k \to \infty$. On the other hand, it is showed that if $\rho \geq \|{\bf F}_{\text{BB}}^{\textrm{H}}\|_{\text{F}}^{2}$, 
\begin{align}
\label{L_diff_1}
    \mathcal{L}^{(k+1)} - \mathcal{L}^{(k)} \leq (\text{i}) \times \left\| {\bf{\tilde F}}_{\text{RF}}^{(k+1)} - {\bf{\tilde F}}_{\text{RF}}^{(k)}\right\|_{\text{F}}^{2} \leq 0,
\end{align}
where the term $(\text{i})$ is defined in Appendix \ref{monotonicity_proof}. Combining \eqref{L_diff_0} and \eqref{L_diff_1} leads to
\begin{align}
\label{F_tilde_k_k1}
    {\bf{\tilde F}}_{\text{RF}}^{(k+1)} = {\bf{\tilde F}}_{\text{RF}}^{(k)}.
\end{align}
The above equation together with 
\begin{align}
\label{U_F_RF_tilde}
{\bf U} = \frac{1}{\rho}\left({\bf F}_{\text{opt}} - {\bf{\tilde F}}_{\text{RF}}{\bf F}_{\text{BB}} \right){\bf F}_{\text{BB}}^{\textrm{H}}
\end{align}
(which is the result of combining \eqref{Update_F_RF_tilde} and \eqref{Update_U}) yields
\begin{align}
\label{U_k_k1}
    {\bf U}^{(k+1)} = {\bf U}^{(k)}.
\end{align}
Since ${\bf F}_{\text{RF}}$ is calculated based on ${\bf{\tilde{F}}}_{\text{RF}}$ and ${\bf U}$ (see Line 6 in Algorithm \ref{Proposed_Alg}), \eqref{F_tilde_k_k1} and \eqref{U_k_k1} yields
\begin{align}
    {\bf F}_{\text{RF}}^{(k+1)} = {\bf F}_{\text{RF}}^{(k)}.
\end{align}
Further, according to Line 8 in Algorithm \ref{Proposed_Alg}, we have
\begin{align}
    {\bf F}_{\text{RF}}^{(k)} = {\bf{\tilde F}}_{\text{RF}}^{(k)}.
\end{align}

On the other hand, from \eqref{F_tilde_k_k1} we have $\left\| {\bf{\tilde F}}_{\text{RF}}^{(k+1)} \!-\! {\bf{\tilde F}}_{\text{RF}}^{(k)} \!\right\|_{\text{F}} \to 0$ as $k \to \infty$. This leads to the fact that: for any positive $\epsilon$, there always exists an integer $T$ (large enough), such that
\begin{align*}
    & \left\| {\bf{\tilde F}}_{\text{RF}}^{(k_1)} \!-\! {\bf{\tilde F}}_{\text{RF}}^{(k_2)} \!\right\|_{\text{F}}  \nonumber \\
    = & \left\| {\bf{\tilde F}}_{\text{RF}}^{(k_1)} \!\!-\! {\bf{\tilde F}}_{\text{RF}}^{(k_1+1)} \!+\! {\bf{\tilde F}}_{\text{RF}}^{(k_1+1)} \!\!-\! {\bf{\tilde F}}_{\text{RF}}^{(k_1+2)} \!+\! \cdots \!+\! {\bf{\tilde F}}_{\text{RF}}^{(k_2-1)} \!\!-\! {\bf{\tilde F}}_{\text{RF}}^{(k_2)} \!\right\|_{\text{F}} \nonumber \\
    \leq & \left\| {\bf{\tilde F}}_{\text{RF}}^{(k_1)} \!\!-\! {\bf{\tilde F}}_{\text{RF}}^{(k_1+1)} \right\|_{\text{F}} + \left\| {\bf{\tilde F}}_{\text{RF}}^{(k_1+1)} \!\!-\! {\bf{\tilde F}}_{\text{RF}}^{(k_1+2)} \right\|_{\text{F}} + \cdots  \nonumber \\
    & \cdots + \left\| {\bf{\tilde F}}_{\text{RF}}^{(k_2-1)} \!\!-\! {\bf{\tilde F}}_{\text{RF}}^{(k_2)} \right\|_{\text{F}} \nonumber \\
    \leq & ~\! \epsilon  \nonumber
\end{align*}
holds for all $k_1$, $k_2$ $\geq T$ (without loss of generality we assume $k_2 > k_1$ in the above inequalities). This indicates that sequence $\left\{ \! {\bf{\tilde F}}_{\text{RF}}^{(k)} \! \right\}$ is a Cauchy sequence, and thus it converges to a fixed point after a finite number (i.e., $T$) of iterations \cite{Rudin1986}. Similarly, both sequences $\left\{ \! {\bf{U}}^{(k)} \! \right\}$ and $\left\{ \! {\bf{F}}_{\text{RF}}^{(k)} \! \right\}$ are Cauchy sequences and they converge to fixed points after $T$ iterations, thanks to \eqref{U_F_RF_tilde} and Line 6 in Algorithm \ref{Proposed_Alg}. This completes the proof. 

\section{Proof of Theorem \ref{AltOptLSADMM_convergence_theorem}}
\label{AltOptLSADMM_convergence_proof}
Since the proposed AltOpt-LS-ADMM algorithm, i.e., Algorithm \ref{Proposed_Alg}, has unique optimal solutions for both ${\bf F}_{\text{BB}}$ (see \eqref{solution_F_BB}) and ${\bf F}_{\text{RF}}$ (see Theorem \ref{ADMM_convergence_theorem}) at each iteration, we have
\begin{align}
    \| {\bf F}_{\text{opt}} - {\bf F}_{\text{RF}}^{(i + 1)}{\bf F}_{\text{BB}}^{(i + 1)} \|_{\text{F}} \leq & ~\! \| {\bf F}_{\text{opt}} - {\bf F}_{\text{RF}}^{(i)}{\bf F}_{\text{BB}}^{(i + 1)} \|_{\text{F}}  \nonumber  \\
    \leq & ~\! \| {\bf F}_{\text{opt}} - {\bf F}_{\text{RF}}^{(i)}{\bf F}_{\text{BB}}^{(i)} \|_{\text{F}},  \nonumber 
\end{align}
which shows that sequence $\left\{\|{\bf F}_{\text{opt}} - {\bf F}_{\text{RF}}^{(i)}{\bf F}_{\text{BB}}^{(i)} \|_{\text{F}}\right\}$ is monotonically decreasing. On the other hand, it is straightforward to see that $\|{\bf F}_{\text{opt}} - {\bf F}_{\text{RF}}^{(i)}{\bf F}_{\text{BB}}^{(i)} \|_{\text{F}}$ is bounded from below by 0. This indicates that sequence $\left\{\| {\bf F}_{\text{opt}} - {\bf F}_{\text{RF}}^{(i)}{\bf F}_{\text{BB}}^{(i)} \|_{\text{F}} \right\}$ generated by the proposed algorithm converges. This completes the proof.  

\section{Proof of Lemma \ref{monotonicity_lemma}}
\label{monotonicity_proof}
The difference between the augmented Lagrangian function values at two successive iterations is calculated as
\begin{align}
    & \mathcal{L} \! \left({\bf{\tilde F}}_{\text{RF}}^{(k+1)}, {\bf F}_{\text{RF}}^{(k+1)}, {\bf U}^{(k+1)}\right) - \mathcal{L} \! \left({\bf{\tilde F}}_{\text{RF}}^{(k)}, {\bf F}_{\text{RF}}^{(k)}, {\bf U}^{(k)}\right)  \nonumber \\
    = & \left[ \mathcal{L} \! \left({\bf{\tilde F}}_{\text{RF}}^{(k+1)}, {\bf F}_{\text{RF}}^{(k+1)}, {\bf U}^{(k+1)}\right) \!-\! \mathcal{L} \! \left({\bf{\tilde F}}_{\text{RF}}^{(k+1)}, {\bf F}_{\text{RF}}^{(k+1)}, {\bf U}^{(k)}\right) \right] \nonumber \\
    & + \left[ \mathcal{L} \! \left({\bf{\tilde F}}_{\text{RF}}^{(k+1)}, {\bf F}_{\text{RF}}^{(k+1)}, {\bf U}^{(k)}\right) \!-\! \mathcal{L} \! \left({\bf{\tilde F}}_{\text{RF}}^{(k)}, {\bf F}_{\text{RF}}^{(k+1)}, {\bf U}^{(k)}\right) \right] \nonumber \\
    \label{difference_L}
    & + \left[ \mathcal{L} \! \left({\bf{\tilde F}}_{\text{RF}}^{(k)}, {\bf F}_{\text{RF}}^{(k+1)}, {\bf U}^{(k)}\right) \!-\! \mathcal{L} \! \left({\bf{\tilde F}}_{\text{RF}}^{(k)}, {\bf F}_{\text{RF}}^{(k)}, {\bf U}^{(k)}\right) \right].
\end{align}
The three terms in the above three square brackets are respectively calculated as follows. The first term is bounded as
\begingroup
\allowdisplaybreaks
\begin{subequations}
\label{U_k}
\begin{align}
    & \mathcal{L} \! \left({\bf{\tilde F}}_{\text{RF}}^{(k+1)}, {\bf F}_{\text{RF}}^{(k+1)}, {\bf U}^{(k+1)}\right) - \mathcal{L} \! \left({\bf{\tilde F}}_{\text{RF}}^{(k+1)}, {\bf F}_{\text{RF}}^{(k+1)}, {\bf U}^{(k)}\right)  \nonumber  \\
    = & ~\! \frac{\rho}{2} \left( \left\|{\bf{\tilde F}}_{\text{RF}}^{(k+1)} - {\bf F}_{\text{RF}}^{(k+1)} + {\bf U}^{(k+1)}\right\|_{\text{F}}^{2} - \left\| {\bf U}^{(k+1)} \right\|_{\text{F}}^{2} \right) \nonumber \\
    \label{U_k_a}
    & - \frac{\rho}{2} \left( \left\|{\bf{\tilde F}}_{\text{RF}}^{(k+1)} - {\bf F}_{\text{RF}}^{(k+1)} + {\bf U}^{(k)}\right\|_{\text{F}}^{2} - \left\| {\bf U}^{(k)} \right\|_{\text{F}}^{2} \right) \\
    \label{U_k_b}
    = & ~\! \frac{\rho}{2} \left( \left\|2{\bf U}^{(k+1)} \!-\! {\bf U}^{(k)} \right\|_{\text{F}}^{2} \!-\! 2\left\|{\bf U}^{(k+1)}\right\|_{\text{F}}^{2} \!+\! \left\|{\bf U}^{(k)}\right\|_{\text{F}}^{2} \right) \\
    = & ~\! \rho \left\|{\bf U}^{(k+1)} - {\bf U}^{(k)}\right\|_{\text{F}}^{2} \nonumber  \\
    \label{U_k_c}
    = & ~\! \frac{1}{\rho} \left\| \! \left( \! {\bf F}_{\text{opt}} \!-\! {\bf{\tilde F}}_{\text{RF}}^{(k+1)}{\bf F}_{\text{BB}} \!\! \right) \!\! {\bf F}_{\text{BB}}^{\textrm{H}} \!-\! \left( \! {\bf F}_{\text{opt}} \!-\! {\bf{\tilde F}}_{\text{RF}}^{(k)}{\bf F}_{\text{BB}} \!\! \right) \!\! {\bf F}_{\text{BB}}^{\textrm{H}} \right\|_{\text{F}}^{2}  \\
    = & ~\! \frac{1}{\rho} \left\| \left({\bf{\tilde F}}_{\text{RF}}^{(k)} - {\bf{\tilde F}}_{\text{RF}}^{(k+1)} \right){\bf F}_{\text{BB}}{\bf F}_{\text{BB}}^{\textrm{H}} \right\|_{\text{F}}^{2}  \nonumber  \\
    \label{U_k_d}
    \leq & ~\! \frac{1}{\rho} \left\| {\bf F}_{\text{BB}}{\bf F}_{\text{BB}}^{\textrm{H}} \right\|_{\text{F}}^{2} \left\| {\bf{\tilde F}}_{\text{RF}}^{(k+1)} - {\bf{\tilde F}}_{\text{RF}}^{(k)}\right\|_{\text{F}}^{2},
\end{align}
\end{subequations}
\endgroup
where in \eqref{U_k_a} we used the definition of $\mathcal{L}\Big({\bf{\tilde F}}_{\text{RF}}, {\bf F}_{\text{RF}}, {\bf U}\Big)$; in \eqref{U_k_b} we employed ${\bf{\tilde F}}_{\text{RF}}^{(k+1)} - {\bf F}_{\text{RF}}^{(k+1)} = {\bf U}^{(k+1)} - {\bf U}^{(k)}$ (due to \eqref{Update_U}); in \eqref{U_k_c} we utilized \eqref{U_F_RF_tilde}; in \eqref{U_k_d} we used the fact that $\|{\bf M}{\bf N}\|_{\text{F}} \leq \|{\bf M}\|_{\text{F}}\|{\bf N}\|_{\text{F}}$ holds for any matrices ${\bf M}$ and ${\bf N}$ of appropriate sizes. The second term is bounded as
\begingroup
\allowdisplaybreaks
\begin{subequations}
\label{F_tilde_k}
\begin{align}
    & \mathcal{L} \! \left({\bf{\tilde F}}_{\text{RF}}^{(k+1)}, {\bf F}_{\text{RF}}^{(k+1)}, {\bf U}^{(k)}\right) - \mathcal{L} \! \left({\bf{\tilde F}}_{\text{RF}}^{(k)}, {\bf F}_{\text{RF}}^{(k+1)}, {\bf U}^{(k)}\right) \nonumber \\
    \leq & ~\! \Re \! \left\{ \! \Big\langle \nabla_{{\bf{\tilde F}}_{\text{RF}}} \mathcal{L} \! \left( {\bf{\tilde F}}_{\text{RF}}^{(k+1)}, {\bf F}_{\text{RF}}^{(k+1)}, {\bf U}^{(k)} \right) , {\bf{\tilde F}}_{\text{RF}}^{(k+1)} \!-\! {\bf{\tilde F}}_{\text{RF}}^{(k)}\Big\rangle \! \right\} \nonumber \\
    \label{F_tilde_k_b}
    & - \frac{\gamma}{2} \left\| {\bf{\tilde F}}_{\text{RF}}^{(k+1)} - {\bf{\tilde F}}_{\text{RF}}^{(k)} \right\|_{\text{F}}^{2}  \\
    \label{F_tilde_k_c}
    = & ~\! - \frac{\lambda_{\textrm{min}}({\bf F}_{\text{BB}}{\bf F}_{\text{BB}}^{\textrm{H}}) + \rho}{2} \left\| {\bf{\tilde F}}_{\text{RF}}^{(k+1)} - {\bf{\tilde F}}_{\text{RF}}^{(k)} \right\|_{\text{F}}^{2},
\end{align}
\end{subequations}
\endgroup
where in \eqref{F_tilde_k_b} we utilized the strongly convexity of the Lagrangian function $\mathcal{L}\Big({\bf{\tilde F}}_{\text{RF}}, {\bf F}_{\text{RF}}, {\bf U}\Big)$ w.r.t. ${\bf{\tilde F}}_{\text{RF}}$ with parameter $\gamma > 0$ \cite{Ryu2016}; in \eqref{F_tilde_k_c} we adopted the optimality condition of \eqref{Update_F_RF_tilde} and $\gamma = \lambda_{\textrm{min}}({\bf F}_{\text{BB}}{\bf F}_{\text{BB}}^{\textrm{H}}) + \rho$ with $\lambda_{\textrm{min}}(\cdot)$ being the minimal eigenvalue of its argument (which is due to the facts that $\mathcal{L}\Big({\bf{\tilde F}}_{\text{RF}}, {\bf F}_{\text{RF}}, {\bf U}\Big)$ is twice continuously differentiable w.r.t. ${\bf{\tilde F}}_{\text{RF}}$, and its strong convexity parameter $\gamma$ satisfies $\nabla_{{\bf{\tilde F}}_{\text{RF}}}^{2} \mathcal{L} = {\bf F}_{\text{BB}}{\bf F}_{\text{BB}}^{\textrm{H}} + \rho{\bf I} \succeq \gamma {\bf I}$ for all ${\bf{\tilde F}}_{\text{RF}}$ \cite{Ryu2016}). Finally, the third term is bounded as
\begingroup
\allowdisplaybreaks
\begin{align}
\label{F_k}
    \mathcal{L} \! \left(\!{\bf{\tilde F}}_{\text{RF}}^{(k)}, {\bf F}_{\text{RF}}^{(k+1)}, {\bf U}^{(k)}\!\right) \!-\! \mathcal{L} \! \left(\!{\bf{\tilde F}}_{\text{RF}}^{(k)}, {\bf F}_{\text{RF}}^{(k)}, {\bf U}^{(k)}\!\right) \leq 0,
\end{align}
\endgroup
where we employed the fact that ${\bf F}_{\text{RF}}^{(k+1)}$ is the minimum of $\mathcal{L}\Big({\bf{\tilde F}}_{\text{RF}}^{(k)}, {\bf F}_{\text{RF}}, {\bf U}^{(k)}\Big)$ according to \eqref{Update_F_RF}.

Substituting the results of \eqref{U_k}, \eqref{F_tilde_k}, and \eqref{F_k} in \eqref{difference_L} yields
\begingroup
\allowdisplaybreaks
\begin{align*}
    & \mathcal{L} \! \left({\bf{\tilde F}}_{\text{RF}}^{(k+1)}, {\bf F}_{\text{RF}}^{(k+1)}, {\bf U}^{(k+1)}\right) - \mathcal{L} \! \left({\bf{\tilde F}}_{\text{RF}}^{(k)}, {\bf F}_{\text{RF}}^{(k)}, {\bf U}^{(k)}\right)  \nonumber \\
    \leq & ~\! \underbrace{\left( \frac{1}{\rho}\left\| {\bf F}_{\text{BB}}{\bf F}_{\text{BB}}^{\textrm{H}} \right\|_{\text{F}}^{2} \!-\! \frac{\lambda_{\textrm{min}}({\bf F}_{\text{BB}}{\bf F}_{\text{BB}}^{\textrm{H}}) \!+\! \rho}{2} \right) }_{\text{(i)}} \left\| {\bf{\tilde F}}_{\text{RF}}^{(k+1)} \!-\! {\bf{\tilde F}}_{\text{RF}}^{(k)}\right\|_{\text{F}}^{2}.
\end{align*}
\endgroup
If $\rho \geq \sqrt{2}\|{\bf F}_{\text{BB}}{\bf F}_{\text{BB}}^{\textrm{H}} \|_{\text{F}}$, the term (i) satisfies: $(\text{i}) \leq 0$, and thus
\begingroup
\allowdisplaybreaks
\begin{align*}
    & \mathcal{L} \! \left({\bf{\tilde F}}_{\text{RF}}^{(k+1)}, {\bf F}_{\text{RF}}^{(k+1)}, {\bf U}^{(k+1)}\right) - \mathcal{L} \! \left({\bf{\tilde F}}_{\text{RF}}^{(k)}, {\bf F}_{\text{RF}}^{(k)}, {\bf U}^{(k)}\right) \leq 0.
\end{align*}
\endgroup
This completes the proof of Lemma \ref{monotonicity_lemma}.

\section{Proof of Lemma \ref{boundedness_lemma}}
\label{boundedness_proof}
By using \eqref{U_F_RF_tilde}, we have
\begingroup
\allowdisplaybreaks
\begin{align}
    & \mathcal{L}\Big({\bf{\tilde F}}_{\text{RF}}, {\bf F}_{\text{RF}}, {\bf U}\Big) \nonumber \\
    = & ~\! \frac{1}{2}\|{\bf F}_{\text{opt}} \!-\! {\bf{\tilde F}}_{\text{RF}}{\bf F}_{\text{BB}}\|_{\text{F}}^{2} \!+\! \frac{\rho}{2} \|{\bf{\tilde F}}_{\text{RF}} \!-\! {\bf F}_{\text{RF}} \!+\! {\bf U}\|_{\text{F}}^{2} \nonumber \\ 
    & - \frac{\rho}{2}\left\|\frac{1}{\rho}\left({\bf F}_{\text{opt}} - {\bf{\tilde F}}_{\text{RF}}{\bf F}_{\text{BB}} \right){\bf F}_{\text{BB}}^{\textrm{H}}\right\|_{\text{F}}^{2}  \nonumber \\
    \geq & ~\! \frac{1}{2}\|{\bf F}_{\text{opt}} \!-\! {\bf{\tilde F}}_{\text{RF}}{\bf F}_{\text{BB}}\|_{\text{F}}^{2} \!+\! \frac{\rho}{2} \|{\bf{\tilde F}}_{\text{RF}} \!-\! {\bf F}_{\text{RF}} \!+\! {\bf U}\|_{\text{F}}^{2} \nonumber \\ 
    & - \frac{1}{2\rho}\left\|{\bf F}_{\text{opt}} - {\bf{\tilde F}}_{\text{RF}}{\bf F}_{\text{BB}} \right\|_{\text{F}}^{2} \left\|{\bf F}_{\text{BB}}\right\|_{\text{F}}^{2}  \nonumber \\
    = & ~\! \frac{1}{2} \left(1 - \frac{1}{\rho}\left\|{\bf F}_{\text{BB}}\right\|_{\text{F}}^{2}\right) \|{\bf F}_{\text{opt}} \!-\! {\bf{\tilde F}}_{\text{RF}}{\bf F}_{\text{BB}}\|_{\text{F}}^{2} \nonumber \\
    & ~\! + \frac{\rho}{2} \|{\bf{\tilde F}}_{\text{RF}} - {\bf F}_{\text{RF}} + {\bf U}\|_{\text{F}}^{2}. \nonumber 
\end{align}
\endgroup
If $\rho \geq \|{\bf F}_{\text{BB}}\|_{\text{F}}^{2}$, then $\mathcal{L}\Big({\bf{\tilde F}}_{\text{RF}}, {\bf F}_{\text{RF}}, {\bf U}\Big) \geq 0$, which completes the proof of Lemma \ref{boundedness_lemma}.

\balance
\bibliographystyle{myIEEEtran}       
\bibliography{refs}

\end{document}